\documentclass[final,5p,times,twocolumn]{elsarticle}
\usepackage{hyperref}

\journal{Chaos, Solitons \& Fractals}
\usepackage{color}

\begin{document}
\begin{frontmatter}

\title{Lessons from being challenged by COVID-19}

\author{E. Tagliazucchi$^1$, P. Balenzuela$^1$, M. Travizano$^2$, G.B. Mindlin$^1$, and P.D. Mininni$^1$}
\address{$^1$Universidad de Buenos Aires, Facultad de Ciencias Exactas y Naturales, Departamento de F\'\i sica, \& IFIBA, CONICET, Ciudad Universitaria, Buenos Aires 1428, Argentina.\\
$^2$ Grandata Labs, 550 15th Street, San Francisco, 94103, California, USA.}

\begin{abstract}
We present results of different approaches to model the evolution of the COVID-19 epidemic in Argentina, with a special focus on the megacity conformed by the city of Buenos Aires and its metropolitan area, including a total of 41 districts with over 13 million inhabitants. We first highlight the relevance of interpreting the early stage of the epidemic in light of incoming infectious travelers from abroad. Next, we critically evaluate certain proposed solutions to contain the epidemic based on instantaneous modifications of the reproductive number. Finally, we build increasingly complex and realistic models, ranging from simple homogeneous models used to estimate local reproduction numbers, to fully coupled inhomogeneous (deterministic or stochastic) models incorporating mobility estimates from cell phone location data. The models are capable of producing forecasts highly consistent with the official number of cases with minimal parameter fitting and fine-tuning. We discuss the strengths and limitations of the proposed models, focusing on the validity of different necessary first approximations, and caution future modeling efforts to exercise great care in the interpretation of long-term forecasts, and in the adoption of non-pharmaceutical interventions backed by numerical simulations.
\end{abstract}

\end{frontmatter}


\section{Introduction}

Since the sudden outbreak of the COVID-19 virus in Wuhan, China, in December 2019, the SARS-CoV-2 epidemic (i.e. the severe acute respiratory syndrome caused by the virus) was declared a pandemic by the World Health Organization on March 11th 2020 and changed the way we live almost all over the world. As of May 2020, around 4 million cases have been detected worldwide, with over 284,000 reported deaths. Great efforts are currently underway towards the characterization of the virus and the treatment of its disease. Without the possibility of a vaccine envisioned for the near future, disease containment has focused mainly on non-pharmaceutical interventions (NPIs) aimed at restricting the circulation of the population, as well as at reducing the risk of contagion in groups of vulnerable individuals. The choice of containment strategy is a key factor to forecast the local severity of the epidemic, as evidenced by the outcome of divergent public health policies adapted by different countries. It is becoming increasingly clear that the development of models to assess the outcome of alternative NPIs will play an important role in the formulation of new public health policies.

\emph{Determination of characteristic disease parameters.} Important first contributions were carried out by studying populations for which the disease spread autonomously, allowing to determine the characteristic parameters of the disease, such as Ref.~\cite{EarlyLombardia}, which allowed the estimation of the effective contagion rate and serial interval in Lombardy, Italy, or the estimation of the infection and case fatality ratios using data from passengers of the Diamond Princess cruise \cite{DiamondPrincess}.

\emph{Previously developed models for seasonal influenza.} SARS-CoV-2 presents multiple similarities and differences with diseases caused by influenza virus that may result in respiratory syndromes. It is then reasonable to build upon previous knowledge when attempting to forecast the evolution of the pandemic. Previous relevant work includes Refs.~\cite{Ferguson2006} and \cite{Halloran2008}, where data-driven agent based models were developed for modeling the spread of seasonal influenza.  In  \cite{VespignaniGleam2010} the authors developed a structured meta-population scheme integrating a stochastic model for disease dynamics, with high-resolution worldwide census data and human mobility patterns at the global scale. In  \cite{Arenas2018}, the authors proposed a theoretical framework for the study of spreading processes in structured meta-populations with heterogeneous agents, subject to different recurrent mobility patterns. All these examples highlight the need for a data-driven approach capable of generating effective descriptions of complex contact patterns \cite{VespingnaniFebrero2020}, and for the development of methods allowing the reuse of empirical contact matrices estimated from different demographics \cite{YMoreno2018}.

\emph{Some basic models adapted for COVID-19.} Models specific for the COVID-19 pandemic have been developed using well-known compartmental models as a starting point. For instance, in \cite{Kuniya2020} the author developed a model based on ordinary differential equations (ODEs) for compartments of susceptible ($S$), exposed ($E$), infected ($I$) and recovered or removed ($R$) individuals (i.e. SEIR model) with the objective of predicting the epidemic peak in Japan. Using this method, the author found a basic reproduction number of $R_0 \approx 2.6$ by adjusting the infection rate to reproduce the accumulated number of cases reported during the first 40 days of the epidemic in Japan.  As another example, in \cite{Brockmann2020} the authors developed a variant of the SEIR model without a compartment for exposed individuals (a SIR model) to fit data obtained from different regions of China, also estimating the local basic reproduction number.

\emph{Regional models for the COVID-19 pandemic.} Several groups have pursued model-based analyses focused on their own countries and regions. For instance, in \cite{Chile2020} the authors analyzed a multi-compartmental SEIR model to estimate the maximal intensive care unit (ICU) bed capacity required in the city of Santiago, Chile, during the peak of the COVID-19 outbreak, while taking different values of the basic reproduction number into account. In \cite{Arenas2020} the authors adapted a meta-population mobility model to capture the spread of COVID-19 in Spain, incorporating empirical epidemiological parameters as well as mobility and census data. The Ferguson group at Imperial College built upon the previously mentioned models for the spread of influenza \cite{Ferguson2006,Halloran2008}, with the purpose of estimating the effect of different potential public health measures, with a focus on Great Britain and the United States of America (USA) \cite{Ferguson1_2020}. 
\emph{Non-pharmaceutical Interventions (NPIs).} One of the most controversial proposed NPI is the formulation of periodic quarantines, which in Ferguson's report are regulated by the availability and saturation of ICU beds. This idea was taken up in \cite{UriAlon2020}, where the authors proposed a cyclic schedule of 4-day work and 10-day lockdown, based on the predictions of a SEIR model with an \emph{ad-hoc} square function modulation of the time dependent reproduction number $R(t)$.

Other  models were developed to study the effect of NPIs on smaller populations, as in the cases of studies performed using data from the Boston residential area in the US. In \cite{YMorenoReport2020} the authors used a data-driven SEIR model (previously developed in \cite{YMoreno2018}) to test six different social distancing strategies, namely (i) school closures, (ii) self-distancing and teleworking, (iii) self-distancing, teleworking, and school closure, (iv) restaurants, nightlife, and closures of cultural venues, (v) non-essential workplace closures, and (vi) total confinement. In \cite{YMorenoVespignaniReport2020}, authors integrated highly detailed  mobility data from cell phone devices, together with census and demographic data, with the purpose of building a detailed agent-based model to describe the transmission dynamics of SARS-CoV-2 in the Boston metropolitan area. The model intended to explore strategies based on lifting  social distancing interventions in combination with testing and isolation of confirmed new cases, contact tracing and quarantining of exposed contacts. A similar study was carried out in \cite{Prem2020}, simulating the outbreak in Wuhan using a deterministic age-structured SEIR model over a 1 year period.

The effects of NPIs were also studied in deterministic models, such as those proposed in \cite{Giordano2020}, where the authors presented a new mean-field epidemiological model for the COVID-19 epidemic in Italy that extended the classical SIR model, called the SIDARTHE model. The model contributed towards evaluating and predicting the effects of implementing different guidelines and protocols (for example, more extensive screening for new cases, or stricter social-distancing).

\emph{Estimations of the reproduction number.} The main goal of NPIs is to flatten the curve of total infected individuals, delaying the peak while keeping the number of cases within the capacity of the health system. Different countries have achieved this objective with varying degrees of success, which can at least be partially explained by their choice of NPIs. Thus, it becomes essential to measure the relative success of each case and to incorporate this knowledge into models capable of assisting decision makers in the formulation of new NPIs. As mentioned before, a key parameter that reflects the dynamics of the epidemics is the reproduction number.  However, in \cite{VespignaniRt2018} authors found that the classical concept of the basic reproduction number $R_0$ is untenable in realistic populations, and provides very little understanding of the evolution of the epidemic. This departure from the classical theoretical picture is not due to behavioral factors and other exogenous epidemiological determinants. Rather, it can be simply explained by the (clustered) contact structure of the population. The authors also provided evidence that methodologies aimed at estimating the instantaneous reproduction number $R(t)$ can be used operationally to characterize the correct dynamics of the epidemic from incidence data.

It should be noted that the correct estimation of $R(t)$ also depends on the accurate detection of infected individuals, alongside other secondary hypotheses. In \cite{Li2020}, the authors estimated the contagiousness and proportion of undocumented infectious cases in China during the weeks before and after the shutdown of travel in and out of Wuhan. They combined data from Tencent (a large social media and technology company) with a networked dynamic meta-population model with Bayesian inference to analyze the early spread within China. Recently, in \cite{Ferguson1_2020} the authors used a semi-mechanistic Bayesian hierarchical model to infer the impact of NPIs  across 11 European countries. The authors studied the course of the epidemic by back-calculating infections from observed deaths, and provided confidence intervals for the impact of different NPIs on the reproduction numbers. A similar analysis was performed in \cite{Zhang2020}.

\emph{Reproduction number and mobility.} It is important to note that even though NPIs can have a noticeable effect on the reproduction number $R(t)$, their effects are not instantaneous; instead, they are modulated by the proper characteristic times of the disease. Since $R(t)$ is an emergent quantity resulting from decreased social contact in large groups of individuals, proposals depending on the capacity to enforce sudden changes of the reproduction number (e.g. as in the modulation of infection rate by a periodic square function, as mentioned above) must be evaluated critically. One of the keys points for understanding the spreading of the disease is the variability in the contagion rate and its dependence on the density and movement of people within or between cities and regions. In this context, in \cite{Christakis2020} the authors used cell phone data from approximately 11 million devices to study how the flow of people through the city of Wuhan contributed to dispersing SARS-CoV-2 throughout China. The authors confirmed the efficacy of lockdown measures for decreasing mobility, and showed that the distribution of population outflow from Wuhan accurately predicted the relative frequency and geographic distribution of COVID-19 infections across China.

\emph{Modeling the spread of COVID-19 in Argentina.} Buenos Aires and its suburbs comprise a megacity offering a unique opportunity to investigate how the virus spreads in large, highly-connected, and densely populated areas, and how real-time mobility information from cell phone devices can be used to assist the prediction of new cases with regional specificity. We believe that the results we will present here can be useful for other researchers being challenged with the task of modeling the early stages of the COVID-19 epidemic in very large cities or in similar demographic contexts.

With this motivation, we summarize our attempts to model the early COVID-19 outbreak and its subsequent evolution in Argentina with a special focus on Buenos Aires and its metropolitan area. We caution the reader to adopt a critical stance towards simple epidemiological models. Even when these models are fed with population census data and realistic mobility estimates, they are fundamentally limited in what they can tell us about the mid- to long-term evolution of the epidemic. While the simplest epidemiological models have several limitations of their own, building more complex models does not necessarily improve our understanding of the epidemic, and can certainly  obfuscate the basic limitations that are intrinsic to the very foundations of the models. Here, models are considered only as ways of exploring the possible future courses of the epidemic, without claims of accuracy, let alone of infallibility. Claims concerning the outcomes of public policies based solely on numerical simulations should be approached with utmost care. The situation is both novel and challenging, full of unknowns that cannot be dispelled by comparison with previous epidemics, and several of the proposed NPIs (such as the enforcement of periodic lockdowns, or the pursuit of massive herd immunity) are unheard of in modern public health policy. The cost of implementing such policies could be huge, and failures could be even costlier. When advising public policy makers, it is our position that models, no matter how detailed, only represent one among multiple sources of information, and should supplement the analysis of actual data, the epidemiological situation, and the advice of public health experts.

This report is structured as follows. First, we consider the problem of interpreting the dynamics of the first cases, with special focus on infectious travelers from abroad (Sec.~\ref{sec:travelers}). Next, we ponder on the dangers of solutions based on instantaneous manipulations of $R(t)$ (Sec.~\ref{sec:magic}). We consider focal dynamics and how they may lead towards the homogenization of the epidemics (Sec.~\ref{sec:homogeneous}). Then, after estimating  relevant parameters from fitting the models to local data in Sec.~\ref{sec:parameters}, we consider coupled models (Sec.~\ref{sec:coupled}) and the role of mobility as estimated from cell phone data (Sec.~\ref{sec:mobility}). We incorporate this data to deterministic and stochastic models, discussing their comparative weaknesses and advantages, and use them to forecast the number of new cases based on possible changes to the isolation policy. Finally, in Sec.~\ref{sec:conclusions} we summarize our conclusions.

\section{The interpretation of the dynamics of the first cases: Travelers as inputs \label{sec:travelers}}

We start with very simple models, and discuss how to study the beginning of the epidemic in the context of a virus known to be spread between countries mostly by air travelers. Throughout this work we will consider models for the officially reported number of infected cases. While the number of actual cases could be larger due to undetected cases, we will assume that as long as testing procedures and protocols are not changed, the percentage of detected cases (with respect to the total number of cases) remains approximately constant. Thus, restricting our analysis to the officially reported number of cases will allow us to decrease the number of unknowns in the system.

\begin{figure}
\begin{center}
\includegraphics[width=1\linewidth]{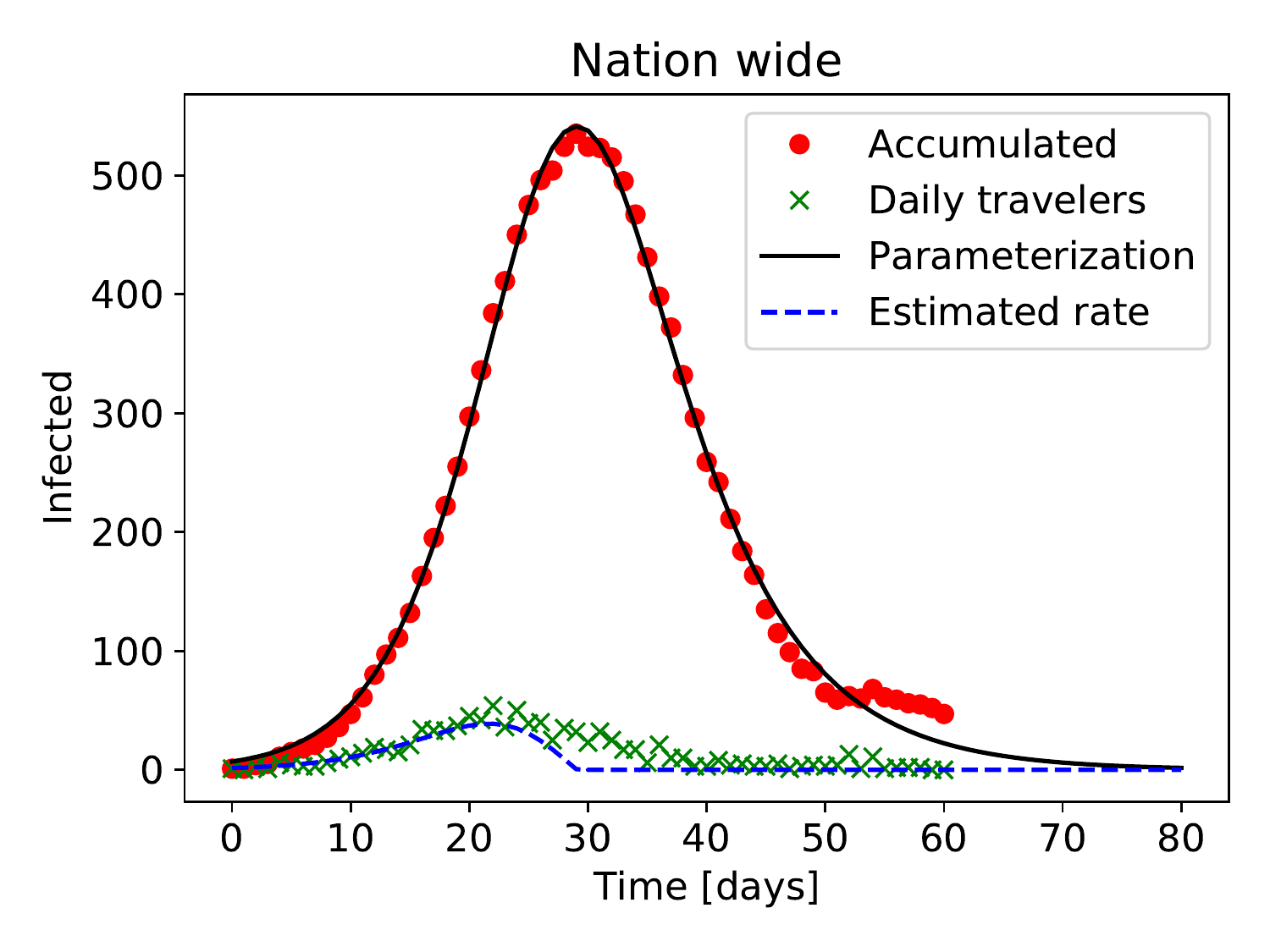}
\includegraphics[width=1\linewidth]{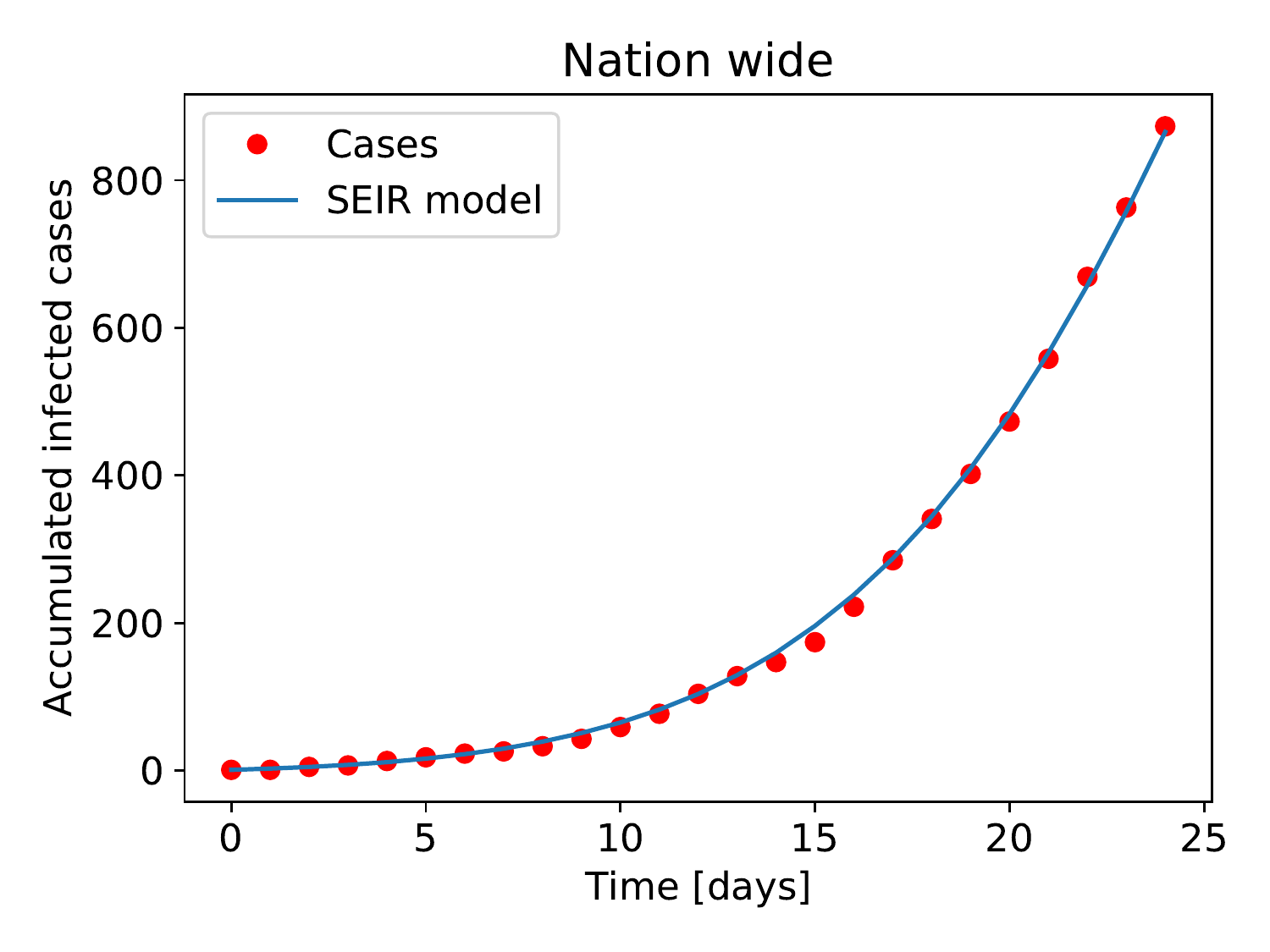}
\end{center}
\caption{\emph{Top:} Evolution of the number of reported infectious travelers arriving to Argentina since the first official case. Dots and crosses correspond to the accumulated (minus recoveries) and new per-day cases, respectively, and the solid and dashed lines represent the best fits to the data. \emph{Bottom:} Number of accumulated cases in the first 25 days, and the best fit obtained using a forced deterministic SEIR model.}
\label{fig:travelers}
\end{figure}

We will progressively move from models that assume homogeneous mixing of the population in the entire country or in large provinces, towards more complex models, including inhomogeneous coupled models. However, we will always assume homogeneous mixing at the level of each individual district. Moreover, we will use cell phone data to reduce the number of unknowns by estimating the coupling between these regions. In our experience, and as will be shown in the following sections, increasing model complexity without incorporating precise empirical information results in models that are hard to validate and interpret, rapidly resulting in homogeneous mixing if ensemble averages are performed to compensate for the lack of detailed knowledge, and thus providing very limited extra information when compared to simpler models.

A brief explanation of Argentinian demographics and the measurements taken by the Argentinian government in the context of the COVID-19 pandemic is in order. Argentina has a total population of nearly 45 millions, distributed between 23 federal states (provinces), plus an autonomous city (the city of Buenos Aires, or CABA). Most of the population lives in this city (2.9 millions) plus its suburbs (the metropolitan area of Buenos Aires, Greater Buenos Aires or AMBA, plus nearby districts, totaling 40 districts with a population over 13 millions). The next most populated regions are the rest of the province of Buenos Aires (PBA), and the provinces of C\'ordoba and Santa Fe. The first officially confirmed case of SARS-CoV-2 was an infected traveler coming from abroad identified on March 3rd. On March 11th, all travellers from countries with confirmed COVID-19 cases were requested to stay in isolation for 14 days. Schools and universities were closed on March 16th, and a complete lockdown was enforced on March 19th. Almost immediately afterwards, borders were closed except for flights bringing stranded Argentinians back from abroad, and vice-versa for foreigners. As a result of these measures, the number of infectious travelers slowly increased until reaching a maximum, and then decreased monotonously with time (see Fig.~\ref{fig:travelers}). The nationwide lockdown was maintained until April 26th, when provinces with zero or very few cases were allowed to relax lockdown rules. In contrast, densely populated areas (over 500,000 inhabitants) or with a large number of confirmed cases were maintained in strict lockdown for at least two more weeks, with the possibility of further extensions to the lockdown.

As mentioned in the introduction, some of the most frequently employed epidemiological models are compartmental; for instance, the SEIR model includes compartments for susceptible ($S$), exposed ($E$), infectious ($I$), and recovered or removed ($R$) individuals. However, since the COVID-19 epidemic started with travelers coming from other countries, modeling the increase of infected individuals solely as an interaction with the compartment of exposed individuals will fail to reproduce the early dynamics of the disease. As discussed next, considering the contribution of these individuals results in more reasonable estimates of the disease transmission rate, which are in better agreement with those obtained in other countries by other methods \cite{DiamondPrincess, Kuniya2020, Brockmann2020, Giordano2020}. We thus briefly summarize how to include incoming infectious individuals in the simplest models, and then adopt the same mechanism when developing more complicated models in the following sections.  

In stochastic models, the number of infectious travelers can be added directly to the infectious compartment. Consider a simple stochastic SEIR model \cite{Lekone}:
\begin{eqnarray}
  S(t+\Delta) &=& S(t) - B(t) , \label{eq:stSEIR_S} \\
  E(t+\Delta) &=& E(t) + B(t) - C(t) , \label{eq:stSEIR_E}  \\
  I(t+\Delta) &=& I(t) + C(t) - D(t) + F(t) , \label{eq:stSEIR_I}
\end{eqnarray}
where $R$ is obtained from fixing the total population $N$ (plus the incoming infectious travelers). Here $B(t) = \textrm{Bin}\left(S(t),P_E(t)\right)$, $C(t) = \textrm{Bin}\left(E(t),P_I(t)\right)$, and $D(t) = \textrm{Bin}\left(I(t),P_R(t)\right)$ are binomial distributions with exponential probability distribution functions $P_E(t) = 1-\textrm{exp}(-\beta \Delta I(t)/N)$, $P_I(t) = 1-\textrm{exp}(-\epsilon \Delta)$, and $P_R(t) = 1-\textrm{exp}(-\gamma \Delta)$, and where $\beta$ is the mean transmission rate, $\epsilon$ is the onset rate (the inverse of the average duration of incubation in days), and $\gamma$ the removal or recovery rate (the inverse of the average duration of the infection). The stochastic variable $F(t)$ represents the number of incoming infectious travelers per day (when $\Delta = 1$ day), and can be obtained from official data.

In deterministic models it is better to fit the data with a smooth function to satisfy mean field approximations. The simplest deterministic and homogeneous SEIR model can be written as
\begin{eqnarray}
  \dot{S} &=& -\beta S I , \label{eq:SEIR_S} \\ 
  \dot{E} &=&  \beta S I - \epsilon E, \\
  \dot{I} &=& -\gamma I  + \epsilon E + F, \\
  \dot{R} &=&  \gamma I .  \label{eq:SEIR_R}
\end{eqnarray}
Note that a forcing term $F$ has been also added to $I$, representing the rate of change in the number of cases from the travelers (i.e., the derivative of $I$ resulting from incoming infectious travelers). We also consider a more complex SEIR model with additional compartments that will be used many times in the rest of this work (see, e.g., \cite{Brockmann2020,Chile2020,Feng}),
\begin{eqnarray}
  \dot{S} &=& -\beta S [J + (1-\rho_I) I + (1-\rho_H) H], \label{eq:SEJIHR_S} \\ 
  \dot{E} &=&  \beta S [J + (1-\rho_I) I + (1-\rho_H) H] - \epsilon E, \\
  \dot{J} &=& -\gamma J + \epsilon (1-\phi) E, \\
  \dot{I} &=& -\gamma I + \epsilon \phi E + F, \\
  \dot{H} &=& -\gamma H + \gamma (1- \chi) I, \\
  \dot{R} &=&  \gamma ( J + \chi I + H), \label{eq:SEJIHR_R}
\end{eqnarray}
Here $J$ is the number of mild infectious individuals, $I$ are moderately infectious individuals, and $H$ are hospitalized individuals. $F$, as before, represents the rate of change in $I$ because of infectious travelers (we assume all detected infectious travelers are moderate, and may be hospitalized later or not). The new coefficients are $\rho_I$ (the fraction of moderate infectious individuals that are properly isolated), $\rho_H$ (the fraction of hospitalized infectious individuals that are isolated), $\phi$ (the fraction of exposed individuals with moderate to severe symptoms) and $\chi$ (the fraction of moderate infectious individuals that do not require hospitalization).

As mentioned before, $F$ should be represented by a smooth function in these ODEs. The product of two logistic functions was found to give a good approximation to the data of accumulated infectious travelers (minus discharged cases, see Fig.~\ref{fig:travelers}), and its derivative caped to only  positive values (as the passage from infectious to recovered compartments is already included in the models) provides a smooth approximation to the daily cases (associated to $F$). Similar results were obtained for each province or district with cases of incoming infectious travelers.

With the simplest SEIR model given by Eqs.~(\ref{eq:SEIR_S})-(\ref{eq:SEIR_R}), we adjust all the data available for Argentina as a homogeneous group with a single local propagation mechanism, but treating separately infectious individuals from abroad. To this end, we fix $\epsilon=1/(5.1 \, \textrm{days})$ and $\gamma=1/(14 \, \textrm{days})$ \cite{Ferguson1_2020, Giordano2020, Brockmann2020}, and we apply a least square approximation to obtain $\beta$ using the method described in \cite{Kuniya2020}. Figure \ref{fig:travelers} illustrates the result of adjusting the model to the number of accumulated cases in Argentina during the first 25 days. The least square approximation yields $\beta \approx 0.22 $ (or, equivalently, a basic reproduction number $R_0 = \beta/\gamma \approx 3$), a value that will be used in the following sections. As the system evolves in time and the epidemic spreads, the role of local infectious cases becomes more relevant, and the forcing due to infectious travelers begins to play a smaller role.

Equipped with this simple model and a first approximation to the relevant parameters, we now consider some NPIs and other proposed solutions based on numerical simulations that have been suggested in the literature.

\section{A caution note on time dependent lockdowns \label{sec:magic}}

It has been  suggested that a way to simultaneously address the need to bound the spread of the pandemic while keeping some significant level of economic activity is the application of time dependent lockdowns. These policies consist in the alternation between periods during which individuals are allowed to move, trade and travel, with others during which social interactions are minimized. Numerical simulations with time dependent parameters have been carried out to model the effect of these policies and indeed, some of these simulations show the suppression of the epidemic \cite{UriAlon2020}.

\begin{figure}
\begin{center}
\includegraphics[width=1\linewidth]{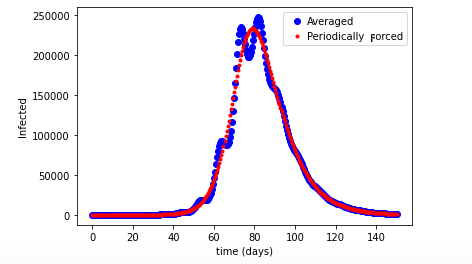}
\end{center}
\caption{Simulation of a deterministic SIR model for a population of $10^6$ individuals. The infection rate parameter is periodically modulated $(\beta_{0}=0.3 \times 10^{-6}$, $A=0.2 \times 10^{-6})$, with $\gamma=0.12$ and $\omega= 2\pi/10$. This corresponds to variations of the basic reproduction number $R_{0} \in [0.8,4.16]$ with a period of 10 days.}
\label{fig:average}
\end{figure}

Yet, it is important to address the viability of these policies by analyzing the general conditions under which suppression could be achieved. Since time dependent lockdowns introduce two time scales (a slow scale characteristic of the epidemic, and a faster scale characteristic of the time dependent policies) we will analyze the problem under the light of the averaging theorem \cite{GHolmes}. This theorem states that for a (slow) dynamical system, periodically forced at a (high) forcing frequency, 
\begin{equation}
\frac{dx}{dt} = \epsilon f(x,t,\epsilon) ,
\end{equation}
with $x \in R^n$,  $0 \leq \epsilon \ll 1$,  $f: R^n \times R \times R^+ \rightarrow R^n$ is $C^r$, and $f(x,t,\epsilon)=f(x,t+T,\epsilon)$, there is an associated autonomous dynamical system (its averaged system), defined as: 
\begin{equation}
\frac{dy}{dt} = \epsilon \frac{1}{T}  \int_{0}^{T} f(y,t:0) \, dt \doteq \epsilon \overline{f}(y), 
\end{equation}
such that there is a change of coordinates mapping the solutions of the original system into those of its associated averaged autonomous system. In other words, up to first order in $\epsilon$, we have that $|x(t)-y(t)| \sim \epsilon$ (if $|x(0)-y(0)| \sim \epsilon$), on a time scale $t \sim 1/\epsilon$.

The application of this theorem to the problem is pertinent. Suppose several periodic interventions are expected during the evolution of the epidemic. Let us model these with an even simpler dynamical model for an epidemic, the  deterministic SIR model:
\begin{eqnarray}
  \dot{S} &=& -[\beta_{0} + A \cos(\omega t) ] S I , \\ 
  \dot{I} &=&  [\beta_{0} + A \cos(\omega t) ] S I - \gamma I \\
  \dot{R} &=&  \gamma I . 
\end{eqnarray}

In these equations, the policy is translated into the periodic modulation of $\beta$.

In Fig.~\ref{fig:average} we compare the simulation of the periodically modulated problem with the solution of its associated averaged version. In this simulation, we studied an epidemic event lasting 100 days, with a time dependent policy of social distancing with a 10 day period. Note that the averaged system constitutes a good approximation of the original problem. The parameters in our simulation correspond to $R(t) \in [0.8,4.16]$, with and average of $R = 2.5$. It is for this reason that in this example there is an epidemic peak. Suppression can indeed be achieved, but only if the average of the modulation leads to $R < 1$. In the case of Argentina, which implemented a long lockdown at an early stage of the epidemic, $R(t)$ reached a value of $\approx 0.9$ as the most optimisic estimate (other countries reported similar values, see, e.g., \cite{Ferguson2_2020,Giordano2020}). It is therefore very difficult (or simply unrealistic) to implement an alternating policy that translates into average parameters leading to suppression. Moreover, as will be shown next, the periods without lockdown may lead to rapid homogeneization of the infected cases in the real system, leading to even longer times for the extinction of the epidemic.

\section{Homogeneous or stochastic? Dynamics of foci, ensembles, and homogeneization \label{sec:homogeneous}}

\begin{figure*}
\begin{center}
\includegraphics[width=0.9\linewidth]{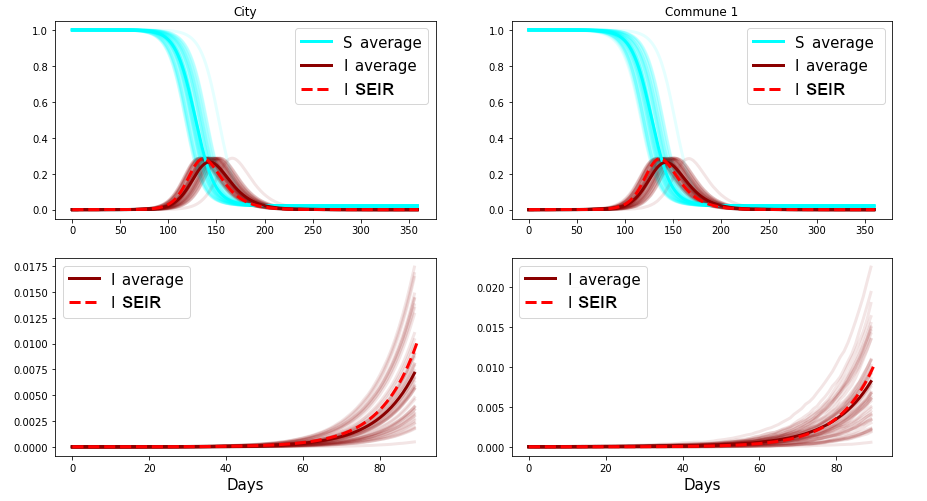}
\end{center}
\caption{Fifty simulations of the stochastic SEIR model for a set of communes inside the city of Buenos Aires (CABA). The top panels display the simulations for the entire city, and for one of the 15 communes. The solutions of the continuous model are displayed, in both cases, with dashed lines. A detail of the simulation at early times is shown in the bottom panels, also for the whole city (left) and for one of the communes (right). The average of the simulations (thick dark lines) is shown together with the solution of a homogeneous deterministic model (dashed lines).}
\label{fig:alhomogeneo}
\end{figure*}

The stochastic modeling of epidemics as in Eqs.~(\ref{eq:stSEIR_S})-(\ref{eq:stSEIR_I}) is important when the number of infectious individuals is small, since in those situations fluctuations can lead to the extinction of the epidemic, even for a highly contagious disease (this can happen, for instance, if the first few infectious individuals had very limited contact with the susceptible population). Typically, an epidemic outbreak starts with few infectious cases that are spatially scattered. If the mobility of those  cases is small, the epidemic will first develop in scattered foci and then slowly spread until all the interconnected susceptible subjects interact with the infectious population, reaching what is known as the homogeneous situation, described, e.g., by the system of Eqs.~(\ref{eq:SEIR_S})-(\ref{eq:SEIR_R}). Typically, the deterministic model used to describe the homogeneous case is therefore thought as the appropriate way to address the evolution of an epidemic outbreak involving a large number of individuals. However, it can also serve as an estimation of the result obtained by analyzing an ensemble of stochastic simulations. To this end, in this section we consider inhomogeneous distributions of infectious cases as obtained by coupling stochastic epidemic models as those in Eqs.~(\ref{eq:stSEIR_S})-(\ref{eq:stSEIR_I}).

To explore the process of homogenization of the epidemic outbreak, we performed an ensemble of stochastic simulations of this system, with a small number of infectious subjects seeded into a susceptible population. We assumed that the population could be grouped into 15 regions (where the number was chosen to reproduce the internal division of the city of Buenos Aires into so-called ``communes", note these are smaller and different from the districts in the suburbs considered later). The number of inhabitants in each commune was obtained from census data. In our numerical experiments, each subject spends half of the day in its commune and a second half either in the same or a different one (the full mathematical details of the coupled stochastic model are described later in Sec.~\ref{sec:coupled}). In the first half of the day, the probability of contagion in the $j^{th}$ commune is given by:
\begin{equation}
P_{E_{j}}(t)=1-exp\left(-\beta \frac{\Delta}{2} I_j(t)/N_j\right) ,
\end{equation}
giving rise to a number of new exposed individuals $E_j(t+\frac{\Delta}{2})=\textrm{Bin}\left(S_j(t),P_{E_j}(t)\right)$. In the second part of the day, we use a mobility matrix \cite{Anapolsky2014} to estimate $m_{i,j}$, the proportion of residents from the $i^{th}$ commune that travel to work in the $j^{th}$ commune. In this way, the number of people that will be found in the $j^{th}$ commune during the second part of the day is $N_j(t+\frac{\Delta}{2})= \sum_{k=1}^{15}m_{k,j}$, leading to $\sum^{15}_{k=1}m_{k,j}I_k(t+\frac{\Delta}{2})/N_{k}$ infectious people. Therefore, in this part of the day, the probability of contagion is:
\begin{equation}
P_{E_j}(t)=1-\textrm{exp}\left[ -\beta \frac{\Delta}{2} \frac{\sum_{k=1}^{15}m_{k,j} I_k(t+\frac{\Delta}{2})}{\sum_{k=1}^{15}m_{k,j}} \right] .
\end{equation}
We can now compute the number $I_{j,i}$ of infectious individuals in the $j^{th}$ commune which are residents of the $i^{th}$ commune as
$I_{j,i}=\textrm{Bin}\left(m_{i,j}S_{i}(t+\frac{\Delta}{2}),P_{E_j}(t+\frac{\Delta}{2})\right)$. Adding all the communes, we obtain,
\begin{equation}
    E_j(t+1)=\sum_k \textrm{Bin}\left(m_{j,k}S_j(t+\frac{\Delta}{2})/N_j,P_{E_k}(t+\frac{\Delta}{2})\right)
\end{equation}
This method of updating stochastic models every half day will be found useful in the following sections, when we use cell phone data to estimate actual mobility between larger districts.

We display 50 simulations in Fig.~\ref{fig:alhomogeneo}, starting with three cases in one commune, with parameters that correspond to $R_0=3.92$ and mobility extracted from \cite{Anapolsky2014}. The dashed line shows the results obtained from integrating the homogeneous model using the parameters that correspond to the whole city. Not surprisingly, the homogeneous continuous model is a good approximation for the large population of Buenos Aires city. Yet, notice how rapidly the dynamics of the city become similar to the dynamics of the first commune. Note again the remarkable similarity between the average of the simulations and the solution of the continuous model in the two bottom panels representing the first days of the stochastic simulations, their average, and the solution of the homogeneous continuous model. 

\begin{figure}
\begin{center}
\includegraphics[width=0.8\linewidth]{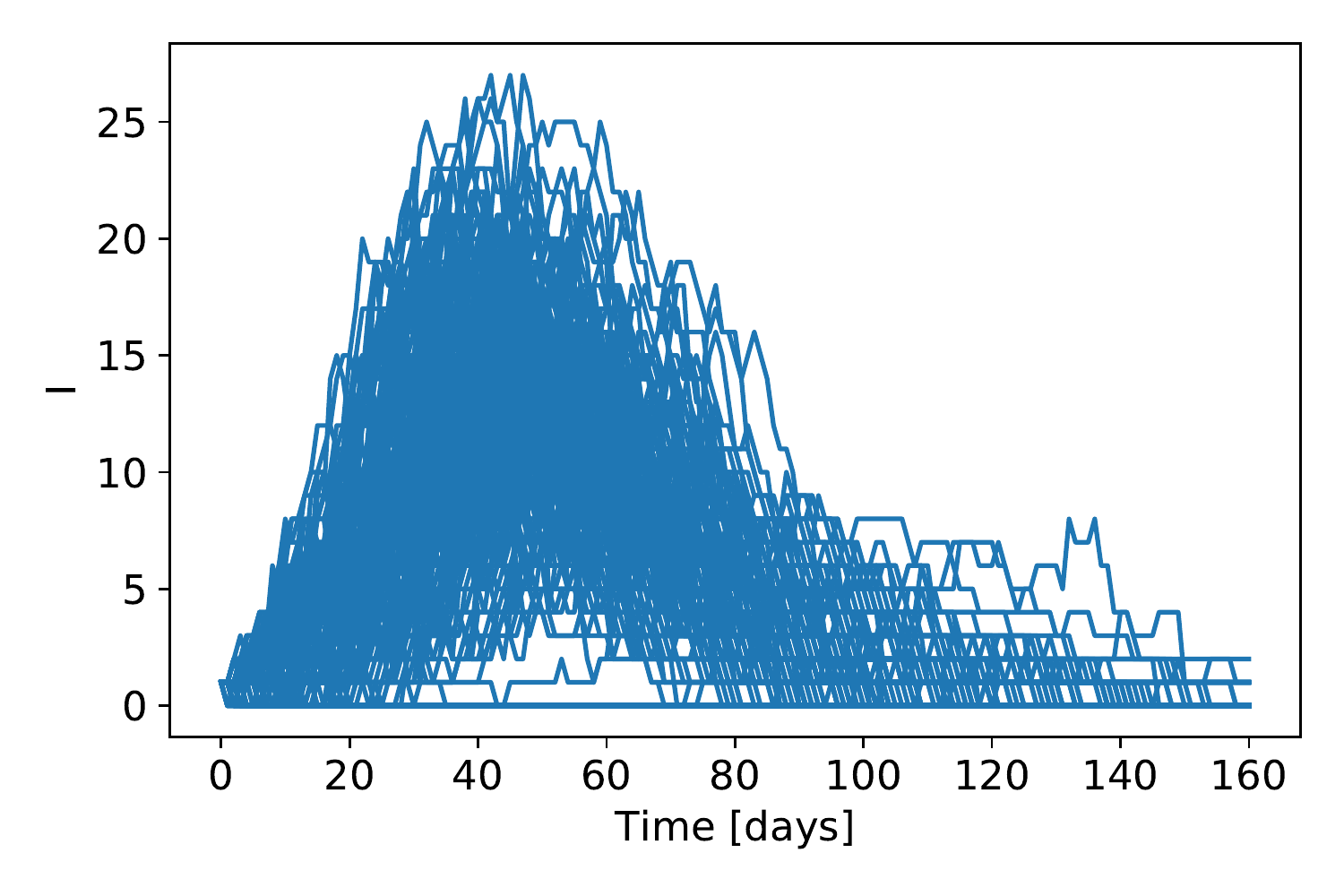}
\includegraphics[width=0.8\linewidth]{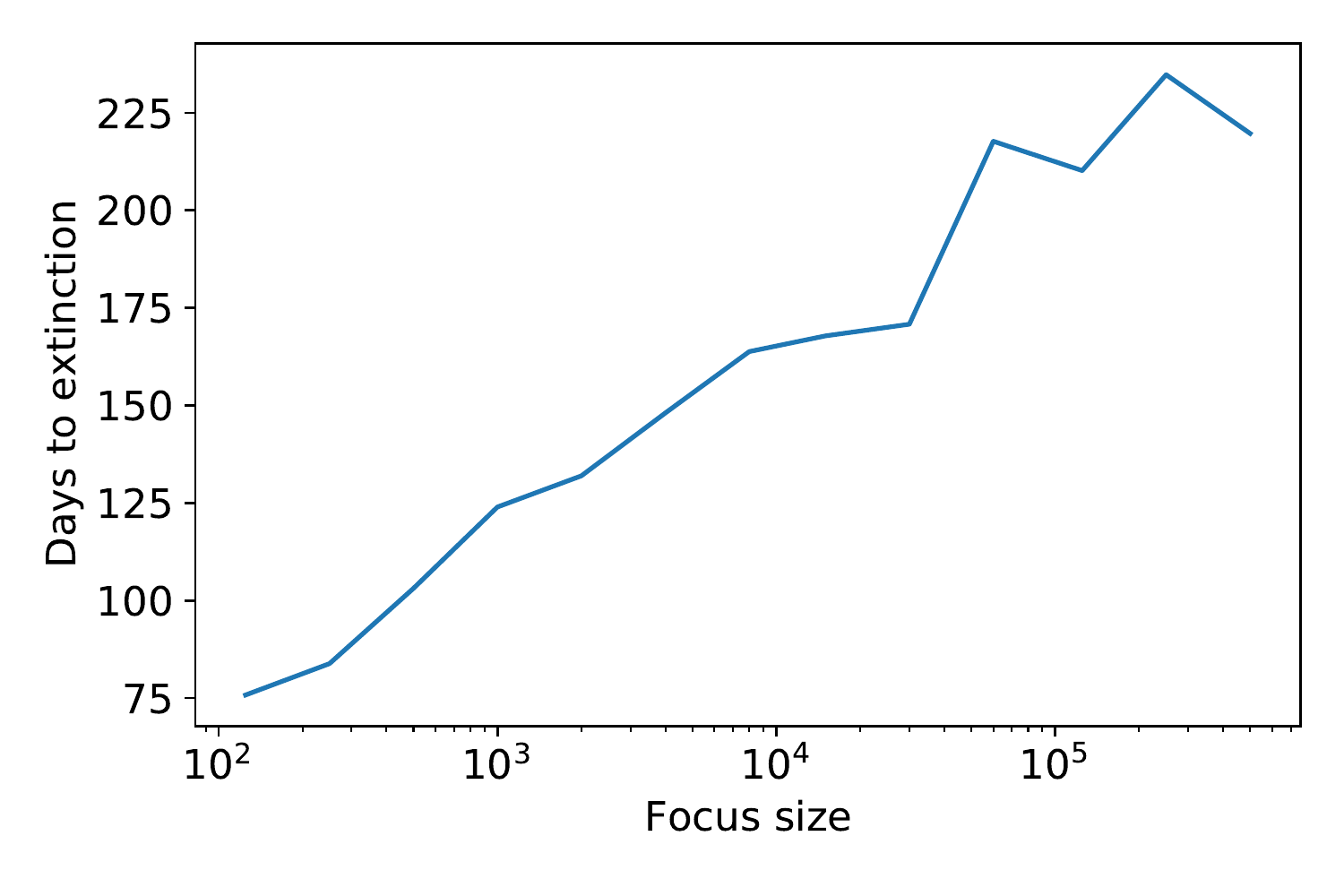}
\end{center}
\caption{\emph{Top:} Evolution of the number of infectious individuals in an ensemble of realizations of a stochastic SEIR model with a population of 50 individuals.  \emph{Bottom:} Time to extinction of the focus as a function of the size of the population involved, assuming the focus is perfectly isolated.}
\label{fig:extinction}
\end{figure}

With these results in mind, we now consider the cost of letting the epidemics spread and evolve towards the homogeneous solution. To this end we consider only one commune, and assume that as a result of the lockdown this commune becomes isolated. Moreover, only a small fraction of the infected population gets in contact with other individuals (i.e., their social network of contacts), resulting in a small focus. Figure \ref{fig:extinction} shows the evolution of the number of infectious individuals in an ensemble of 200 realizations of the stochastic SEIR model with a focus restricted to 50 individuals. The number of infectious individuals quickly drops to zero in some realizations, while in others infectious cases persist for longer times. The mean time for the extinction of the focus can be estimated from such an ensemble. This is also shown in Fig.~\ref{fig:extinction}, which presents this time as a function of the size of the population in the focus. Note the exponential dependence between these quantities: avoiding the growth of the foci becomes crucial for the long-term evolution of the system.

Finally, the results in this section indicate that, to a certain extent, we can approximate the dynamics in the city of Buenos Aires (and probably as well in the individual districts of its metropolitan area) using homogeneous models (see Sec.~\ref{sec:parameters}), and then coupling these districts assuming that the mix between them can be inhomogeneous (see Secs.~\ref{sec:coupled} and \ref{sec:mobility}).

\section{Estimation of transmission rates from local fitting and first evaluation of lockdowns \label{sec:parameters}}

To study the evolution of the epidemics in highly populated regions using locally homogeneous models, and later incorporating coupling to reflect inhomogeneous mix between different regions, we need reasonable estimations of infection rates. These estimations serve the double purpose of avoiding over-fitting in complex models, and evaluating the effect of lockdown measures. Otherwise, conducting ensemble averages over multiple realizations with different parameters would result again in the homogeneization of the results. In this section we show how simple deterministic SEIR models (as the SEJIHR model in Sec.~\ref{sec:travelers}) can be used to estimate these parameters from the data and to study the global and local evolution of the epidemic, assuming weak coupling between regions as a consequence of the lockdown (as will be confirmed later in the analysis of cell phone data). 

First of all, to avoid over-fitting we choose to fix as many parameters as possible based on the literature and available data. We decided to fix all parameter values except for $\beta$. We introduce time dependence, $\beta(t)$, to capture effective changes associated with the lockdown measures. As in previous sections, we fix $\epsilon=1/(5.1 \, \textrm{days})$ and $\gamma=1/(14 \, \textrm{days})$. The other parameters in Eqs.~(\ref{eq:SEJIHR_S})-(\ref{eq:SEJIHR_R}) were chosen as $\rho_I = 0.2$, $\rho_H = 0.9$, $\phi = 0.7$, and $\chi = 0.8$, which yielded a fraction of mild, moderate, and hospitalized infectious individuals compatible with official reports \cite{MinSal}.

\begin{figure}
\begin{center}
\includegraphics[width=1\linewidth]{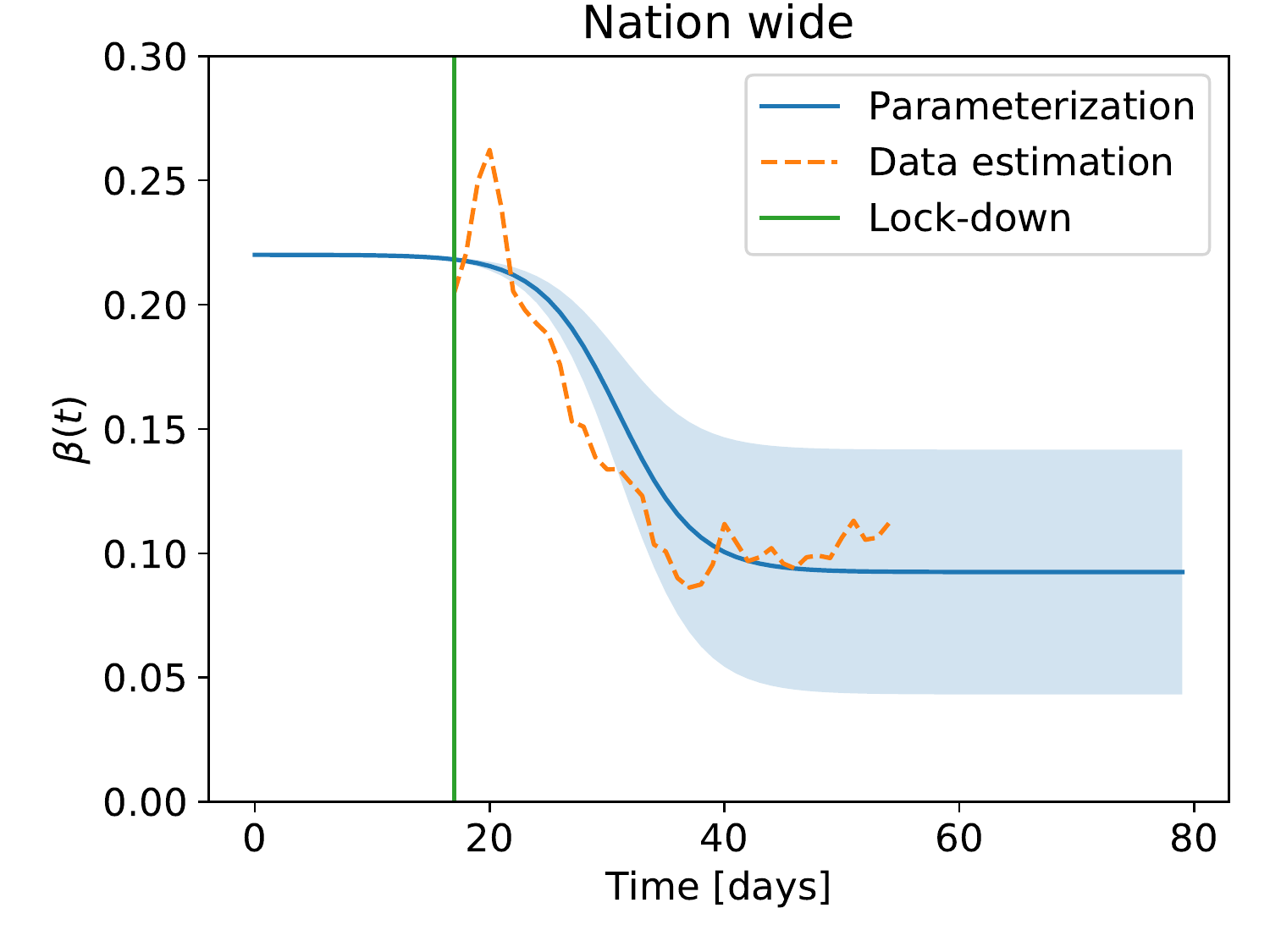}
\includegraphics[width=1\linewidth]{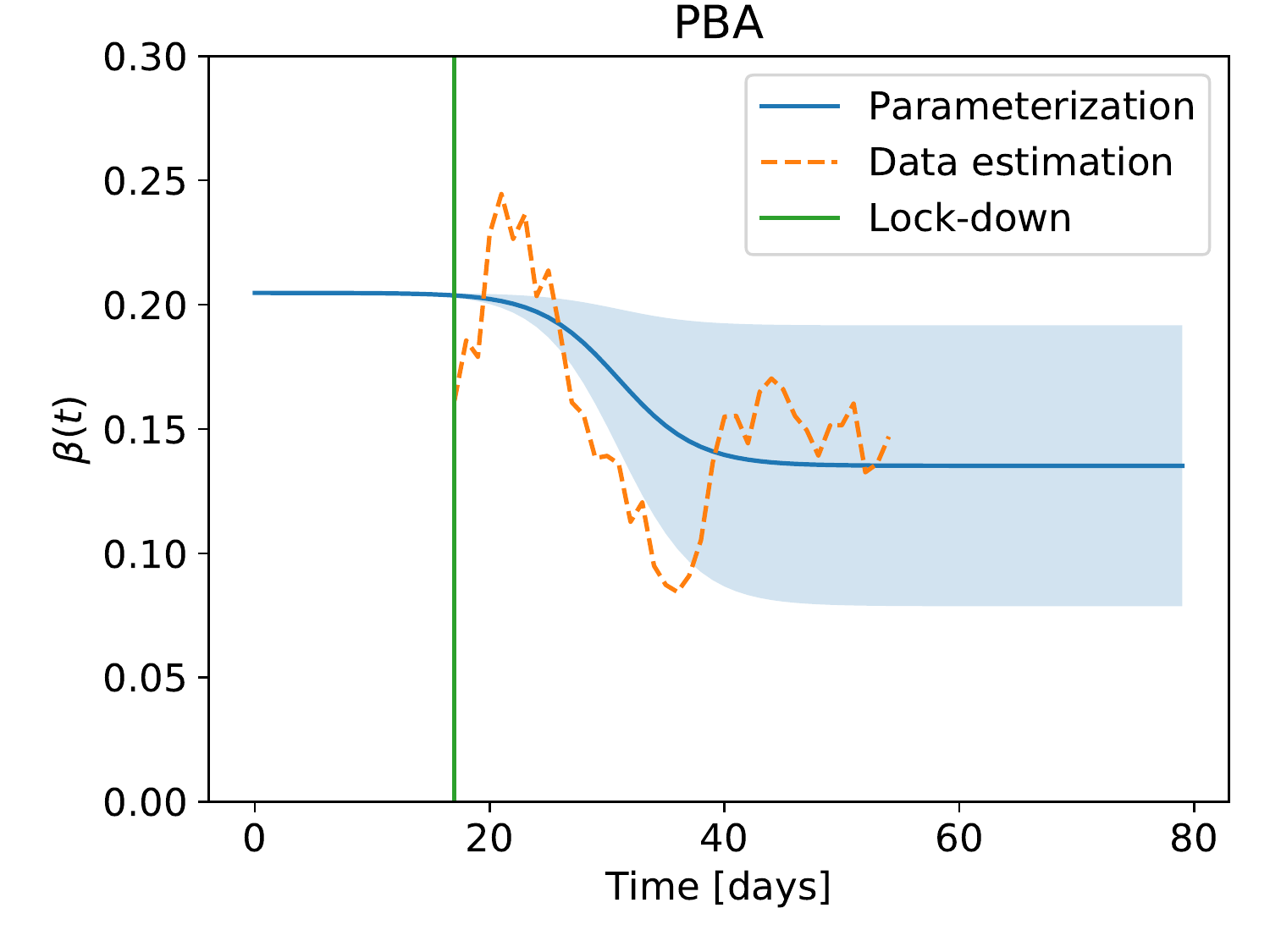}
\end{center}
\caption{Estimated transmission rate $\beta(t)$ for the whole nation (Nation wide) and for the province of Buenos Aires (PBA). The dashed line corresponds to the values obtained from a least square fit with the deterministic model using a moving window of 10 days. The blue line corresponds to the smooth estimation at early and late times, while the blue shading indicates a window of confidence of $95\%$ for $\beta(t)$ during the lockdown.}
\label{fig:beta}
\end{figure}

Then, we fitted the ODEs to the first 25 days of data (i.e. up to 8 days after the beginning of the lockdown) to estimate $\beta_0=\beta(t=0)$, and from the 10th day of the lockdown to the last day with available official information to estimate the mean $\beta(t)$ during the lockdown. These two values were then interpolated with a logistic function to obtain a smooth version of $\beta(t)$ for the model (as will be shown later from the analysis of cell phone data, this is in agreement with observations that the mobility of individuals decreases slowly in time, starting even before the lockdown is put in place). To produce forecasts using the model, it is necessary to calculate the errors that occur when estimating $\beta(t)$ in this way. We thus also fitted the model to the data using least squares with a moving window of 10 days to estimate the dispersion $\sigma_\beta$ in the values of $\beta(t)$. Results for the whole nation and for the most populated province in the country are shown in Fig.~\ref{fig:beta}.

\begin{figure}
\begin{center}
\includegraphics[width=1\linewidth]{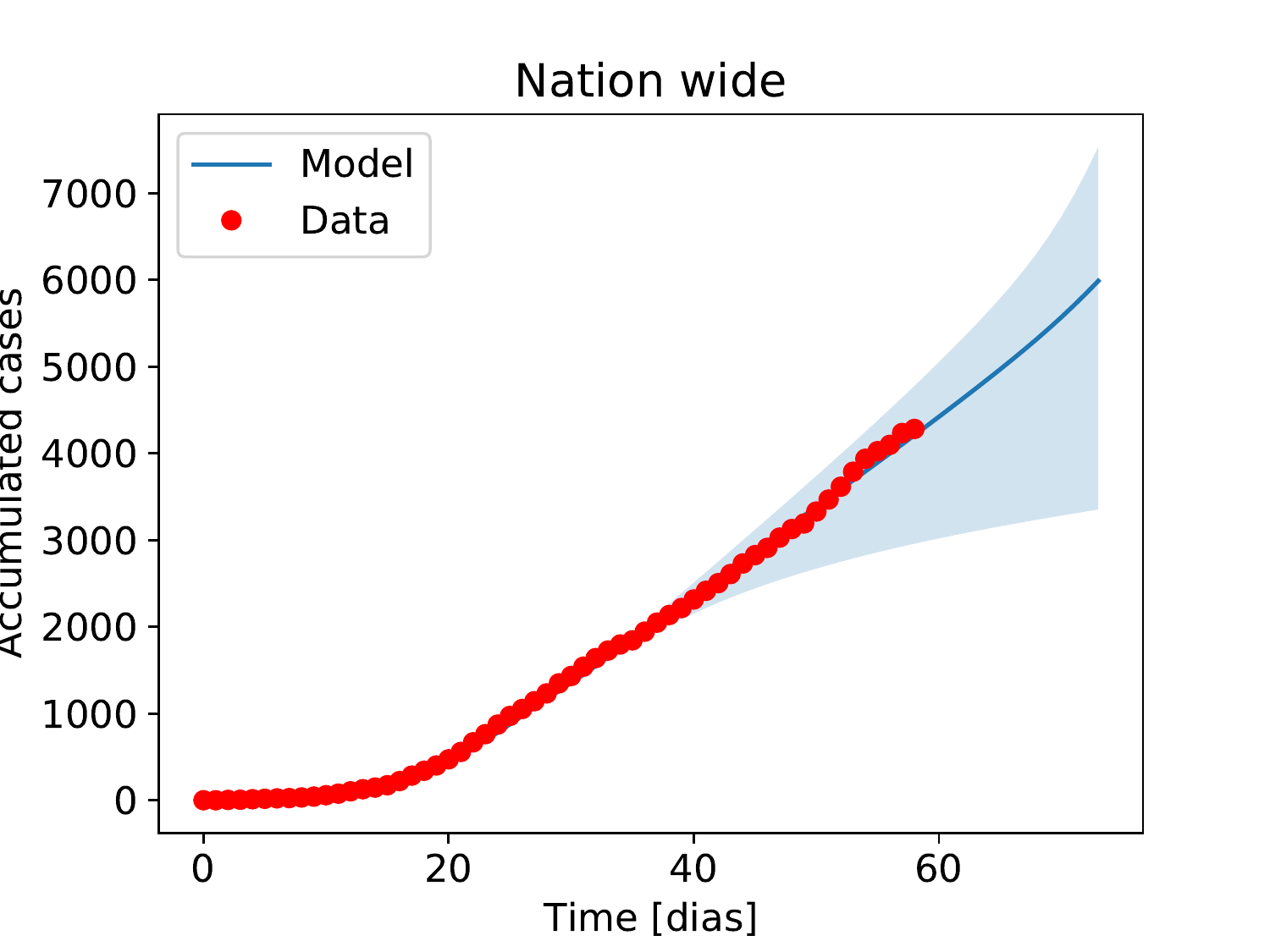}
\includegraphics[width=1\linewidth]{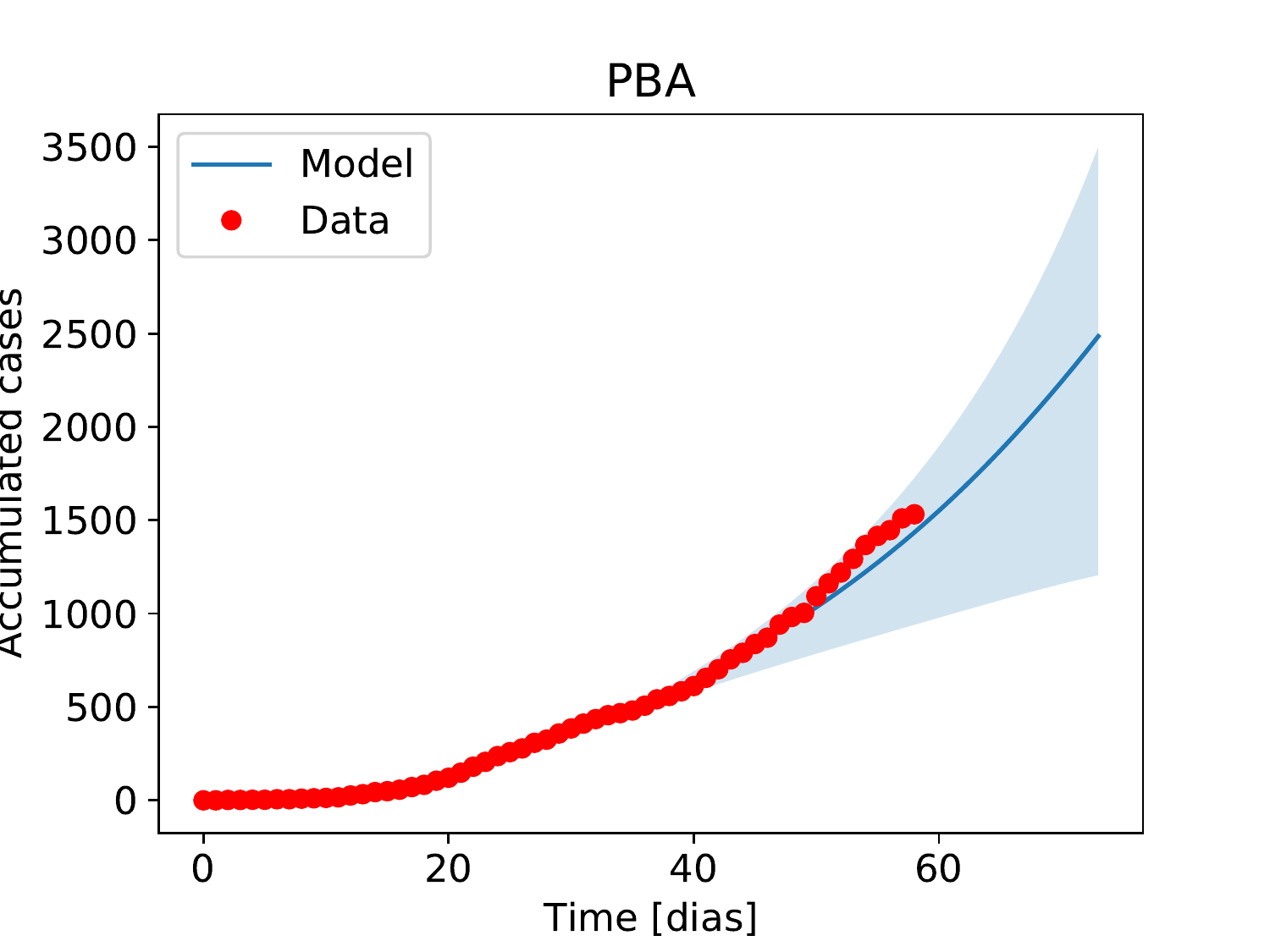}
\end{center}
\caption{Accumulated cases in the entire country (Nation wide) and in the province of Buenos Aires (PBA, {\emph right}), obtained from the SEJIHR model using $\beta(t)$, with 95\% confidence intervals (blue shading) and compared with the official data.}
\label{fig:window}
\end{figure}

When forecasting the possible future outcomes of the epidemic, it is necessary to carry out a series of simulations with different parameter choices. Integration of the system with $\beta(t)\pm 1.96 \, \sigma_\beta$ allowed us to compute solutions within 95\% confidence levels. Results are shown in Fig.~\ref{fig:window}. While  the best fit to the data indicates a decrease in the effective value of $\beta(t)$ as a result of the lockdown when computed for the entire country (with a slight increase at later times), in the province of Buenos Aires (PBA) the decrease in the value of $\beta(t)$ is smaller and presents larger fluctuations (resulting in a different increase in the number of cases in Fig.~\ref{fig:window}). This indicates the need of factoring individual mobility in the analysis, which will be shown next to correlate with the observed behavior of $\beta(t)$. Also, note that the uncertainty of solutions within 95\% confidence intervals increases very rapidly, thus forbidding reliable long-term forecasts of the number of cases. This behavior is to be expected in systems with solutions that grow exponentially at early times, and provides a warning message for all attempts at long-term forecasting of the pandemic.

Finally, when considering separate districts within PBA, the same qualitative behavior (but with different values of $\beta(t)$) is observed. We conclude that inhomogeneous models with coupled districts are required, at least at this level of description.

\section{Coupled models and the role of mobility \label{sec:coupled}}

\begin{figure*}
\begin{center}
\includegraphics[width=1\linewidth]{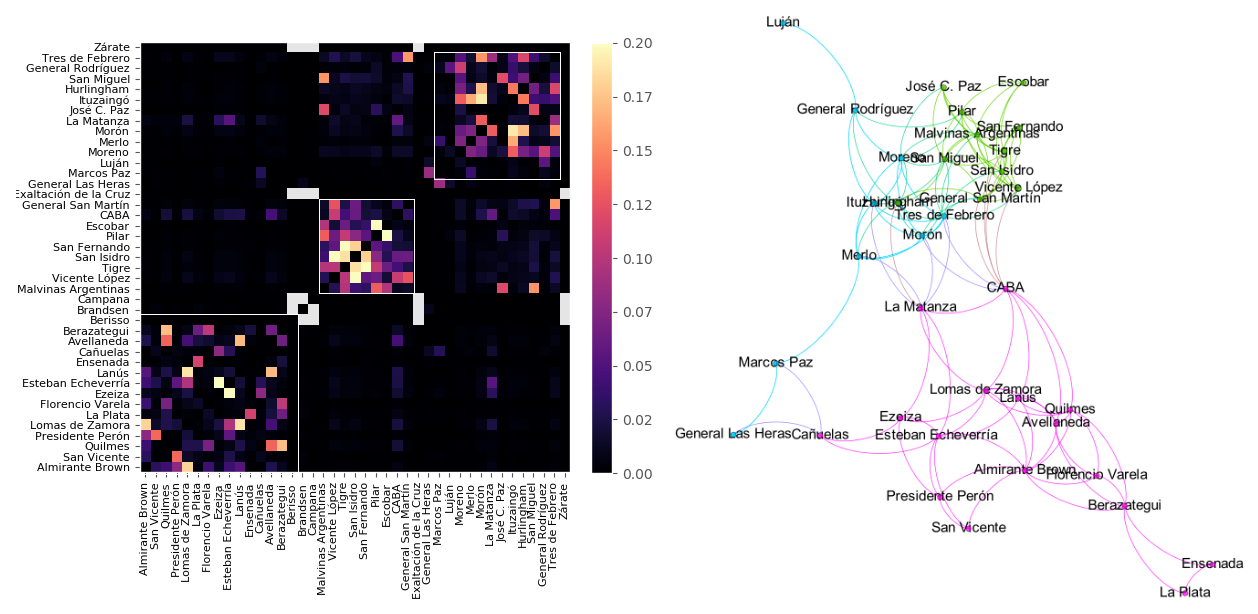}
\end{center}
\caption{\emph{Left:} Normalized inter-regional mobility computed during the 2nd of March. Regions are ordered according to their community membership detected with the Louvain algorithm. The three white diagonal squares indicate major communities corresponding to the southern, centre-northern and western part of the metropolitan area. \emph{Right:} Topological representation of the major communities using the Yifan Hu Multilevel layout algorithm as implemented in Gephi. }
\label{fig:mobility1}
\end{figure*}

As previously done in Sec.~\ref{sec:homogeneous}, both stochastic and deterministic models can be modified to include several interacting geographical regions, each with its own transmission rate $\beta_i$, and number of susceptible ($S_i$), exposed ($E_i$), infectious ($I_i$), and recovered ($R_i$) individuals. We first describe in detail a coupled stochastic model obtained from adapting Eqs.~(\ref{eq:stSEIR_S})-(\ref{eq:stSEIR_I}) to include empirical mobility data. Second, we introduce a coupled deterministic model. Mobility data will be introduced by quantifying a daily mixing of individuals given by $M_{i,j}$ \footnotetext{Note that this matrix represents total numbers of individuals and not proportions, as was the case for $m_{i,j}$ in Sec. \ref{sec:homogeneous}}. The $i,j$ entry of this matrix contains the number of individuals from the $i^{th}$ region who visit region $j^{th}$ during the day, so that $\sum_j{M_{i,j}}=N_i$ is the total population of the $i^{th}$ region, and $\sum_i{M_{i,j}}=N^*_j$ is the number individuals found in the $j^{th}$ region during the day. Coupling is done using the methods described in \cite{Lloyd}. In this section and the following, unlike Sec.~\ref{sec:homogeneous}, regions will correspond to the entire city of Buenos Aires, and to each district in its metropolitan area.

As done before, in the stochastic model we divide each day into two halves of duration $\frac{\Delta}{2}$. During the first half of the day individuals interact according to a mixing matrix $M_{i,j}$, reflecting a typical work and school day (if schools are open). In this case, the exponential probability function for exposure to infected individuals depends on the geographical region and is computed as $P_{E,i}(t) = 1-\textrm{exp}(-\beta_i \frac{\Delta}{2}I_i^*(t)/N_i^*)$ where $I_i^*(t) = \sum_i{M_{i,j}} I_j/N_j$ is the number of infected individuals found in the $i^{th}$ region during the first half of the day. In the same way, we introduce $S_i^*(t) = \sum_i{M_{i,j}}S_j/N_j$ as the number of susceptible individuals in the $i-th$ region during the first half of the day. The probability functions $P_I(t)$ and $P_R(t)$ do not depend on the region since they only contain parameters intrinsic to the disease.

The number of individuals in each compartment per region is then updated as follows,
\begin{eqnarray}
  S_i\left(t+\frac{\Delta}{2}\right) &=& S_i(t) - B_i(t) \label{eq:stSEIR_Si} \\
  E_i\left(t+\frac{\Delta}{2}\right) &=& E_i(t) + B_i(t) - C_i(t) \label{eq:stSEIR_Ei}  \\
  I_i\left(t+\frac{\Delta}{2}\right) &=& I_i(t) + C_i(t) - D_i(t)  \label{eq:stSEIR_Ii}
\end{eqnarray}
with $R_i = N_i - S_i - E_i - I_i$, $B_i(t) = \textrm{Bin}\left(S_i^*(t),P_{E,i}(t)\right)$, $C_i(t) = \textrm{Bin}\left(E_i(t),P_I(t)\right)$ and $D_i(t) = \textrm{Bin}\left(I_i(t),P_R(t)\right)$.

The update rule for the second half of the day can be obtained from the same equations with $M_{i,i} = N_i$ and 0 otherwise, and by adding a forcing term $F_i(t)$ to Eq.~(\ref{eq:stSEIR_Ii}), reflecting the number of incoming infectious travelers for that region.  Note that the transmission rate during the second half of the day, $\beta_i$, is also region-specific. Also, note that both $\beta_i$ and $M_{i,j}$ may depend on time, reflecting the changes in local and inter-regional mobility imposed by quarantine and isolation policies.

\begin{figure*}
\begin{center}
\includegraphics[width=1\linewidth]{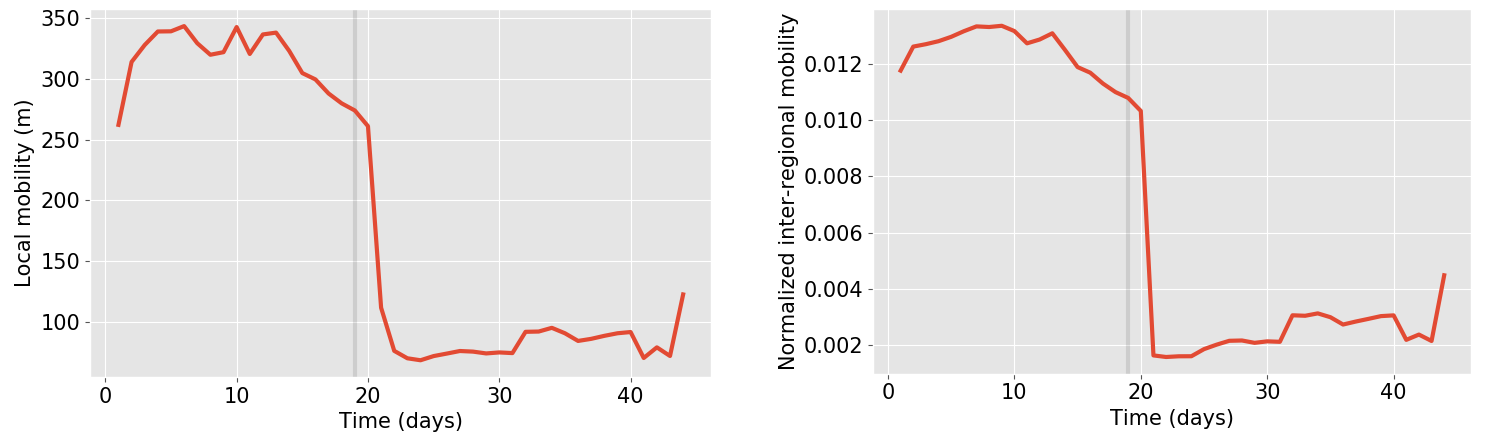}
\end{center}
\caption{\emph{Left:} Average local mobility from the 1st of March to the 12th of April. \emph{Right:} Average inter-regional mobility (computed as the average of all rows in $C_{i,j}$) from the 1st of March to the 12th of April. In both panels the grey vertical line indicates the start of the lockdown (19th March).}
\label{fig:mobility2}
\end{figure*}

Finally, we also consider a coupled version of the homogeneous SEJIHR model introduced in Eqs.~(\ref{eq:SEJIHR_S})-(\ref{eq:SEJIHR_R}). Coupling is introduced by writing equations for $S_i$, $E_i$, $J_i$, $I_i$, $H_i$, and $R_i$, and coupling susceptible and (mild and moderate) infectious individuals in different districts $S_iI_j$ and $S_iJ_j$ (with $i\neq j$) in the equations for $\dot{S_i}$ and $\dot{E_i}$. Thus, the equations for the evolution of $S_i$ and $E_i$ result
\begin{eqnarray}
\dot{S}_i &=& - \beta_i [J_i + (1-\rho_I) I_i] \nonumber \\
  {} && - \sum_{j \neq i} \beta_0 C_{i,j} [J_j + (1-\rho_I) I_j] , \\
\dot{E}_i &=& \beta_i [J_i + (1-\rho_I) I_i] \nonumber \\
  {} && + \sum_{j \neq i} \beta_0 C_{i,j} [J_j + (1-\rho_I) I_j] - \epsilon E_i,
\end{eqnarray}
where $C_{i,j}$ is a normalized and symmetrized version of the mixing matrix given by $C_{i,j} = \frac{M_{i,j} + M_{j,i}}{N_i + N_j}\frac{N_j}{N_i}$ (the last factor $N_j/N_i$ results from taking into account the different populations in each district, and the fact that deterministic models are invariant to the size of the total population so the SEJIHR variables are normalized to 1). We also force each region with their incoming infectious travelers, and thus the equation for $I_i$ is
\begin{equation}
\dot{I}_i = - \gamma I_i + \epsilon \phi E_i + F_i,
\end{equation}
where $F_i$ is a smooth time-dependent function obtained from the data of each district as explained in Sec.~\ref{sec:travelers}. As in the case of the stochastic model, both $\beta_i$ and $C_{i,j}$ potentially depend on time, and $\epsilon$, $\gamma$, $\rho_I$, $\rho_H$, $\phi$, and $\chi$ are time- and region-independent constants chosen as in the previous sections.

\section{Mobility estimation from cell phone location data \label{sec:mobility}}

In this section we describe the use of cell phone location data to estimate the mixing and contact matrices introduced in Sec.~\ref{sec:coupled}, as well as the effective local infection rates, focusing on the advantages and limitations intrinsic to this type of data. 

We obtained geolocated data from anonymized mobile phones spanning the whole country between March 1st and April 12th ($\approx 3 \times 10^6$ unique phones). For each phone we obtained a sequence of latitude/longitude pairs with an associated timestamp, each corresponding to the use of a certain (also anonymized) app (resulting in 100 to 200 events per cell phone per day). Each sequence of latitude/longitude pairs was converted to a sequence of labels indicating the location among the 529 departments in the country, as defined by the National Geographical Institute of Argentina (IGN, \url{https://www.ign.gob.ar/}). As in the previous sections, we restrict our analysis to a subset of 41 departments (or districts) spanning Buenos Aires and the Greater Buenos Aires metropolitan area, comprising approximately $50\%$ of the phones in the dataset.

A major limitation of this type of data is uneven spatial and temporal sampling. Even within the most densely sampled area (Buenos Aires and its metropolitan area) we observed large differences in the number of cell phones relative to the total population. Since for all purposes the temporal sampling of the trajectory must be considered as random, we determined the regional membership of each phone as its most visited region during the day (i.e., the mode of the IGN department labels). 

We computed two metrics per day from this dataset. First, we estimated the mixing matrix $M_{i,j}$ by counting the number of cell phones belonging to the $i^{th}$ region found throughout the day in the $j^{th}$ region, and then rescaling rows so that $\sum_j{M_{i,j}} = N_i$. We note that, since a single cell phone can be found in several regions throughout a day, this mixing matrix represents a coarse temporal average across a time step of $\frac{\Delta}{2}$ in the stochastic coupled model. Better approximations could be eventually obtained dividing each day into shorter intervals. We also computed the normalized and symmetrized contact matrix $C_{i,j}$ as described in the previous section. Figure \ref{fig:mobility1} (left panel) presents a typical contact matrix (computed for the 2nd of March), with regions ordered according to their community membership detected with the Louvain algorithm (resolution parameter $\gamma$ = 1) \cite{blondel2008fast}. Figure \ref{fig:mobility1} (right panel) presents a topological representation of the mixing matrix and its major communities using the Yifan Hu Multilevel layout algorithm \cite{hu2005efficient} as implemented in Gephi (\url{https://gephi.org/}). The three white diagonal squares indicate major communities corresponding to the southern, centre-northern and western part of the metropolitan area, corresponding to purple, green and light blue communities in the right panel, respectively. Next, we estimated $M_i$, the local mobility of region $i$, as the average of all geodesic distances between successive latitude/longitude pairs within that region. As described in the following sections, we use $M_i$ to construct an effective time-dependent and region-specific transmission rate.

Finally, we observed a strong modulation of the local mobility and the inter-regional traffic as a consequence of the lockdown starting on the 19th of March (Figure \ref{fig:mobility2}). Both curves show a gradual decline during the days prior to the beginning of the lockdown, and a very slow recovery of the circulation during the first days of April. 

\subsection{Coupled deterministic models}

For the sake of simplicity, we first consider the deterministic model coupled using cell phone data, and leave the stochastic model (which requires ensemble averages) for the next section. To integrate the model, $\beta_0$ is fixed at the value initially obtained from the best fit to all cases in the country, $\beta_0=0.22$. The temporal evolution of the $\beta_i$ coefficients is approximated as
\begin{equation}
\beta_i(t) = \beta_0 \hat{M_i}(t),
\end{equation}
where $\hat{M_i}(t)$ is a temporally smooth approximation to the local mobility  $M_i(t)$ obtained from the cell phones data, and normalized such that $\langle \hat{M_i}(t) \rangle = 1$, where the brackets denote the time average across the first 14 days (i.e., before the lockdown measures were in place). The coupling between districts $C_{i,j}(t)$ was computed directly from the time-dependent cell phone mobility matrix $M_{i,j}$ using the definition in Sec.~\ref{sec:coupled}. Note that in this way, there are no free parameters left to adjust, resulting in a system whose dynamics can either match the official number of cases or deviate strongly from it, depending on the validity of the parameters, empirical sources of data, and assumptions of the model.

\begin{figure}
\begin{center}
\includegraphics[width=1\linewidth]{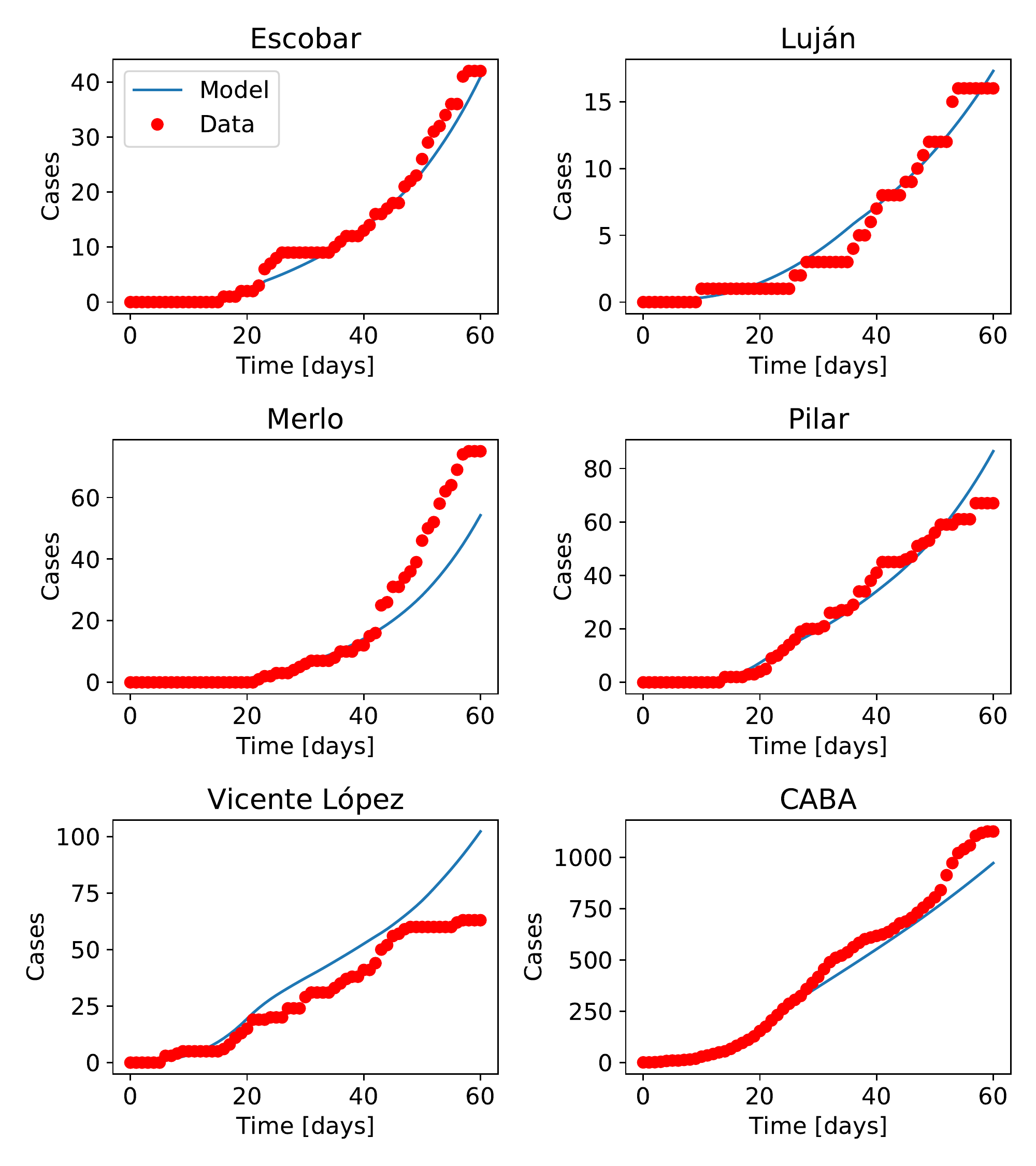}
\end{center}
\caption{Number of accumulated cases in five districts of the Buenos Aires metropolitan region, and in the city of Buenos Aires (CABA). Infectious travelers were reported in the early stages of the outbreak for all these districts. Red dots indicate the official number of cases, while the blue curves indicate results from the deterministic coupled model.}
\label{fig:voyagers}
\end{figure}

We integrated this coupled model for the city of Buenos Aires (CABA) and the 40 districts in the Buenos Aires metropolitan area. This resulted in simulations for 41 districts with very different populations and number of infected individuals. Some districts had less than 10 reported cases at the time when this report was written, while others had several hundreds. Also, some districts had cases of infectious travelers coming from abroad during the first days of the outbreak, while others reported no infectious travelers. Moreover, the first cases in each district were reported at different days. 

We integrated the model for all districts with zero cases as initial condition, and let the incoming travelers drive the increase in the number of infected individuals. We note that in absence of coupling, districts with no incoming infectious travelers would remain at zero cases during the integration time.

\begin{figure}
\begin{center}
\includegraphics[width=1\linewidth]{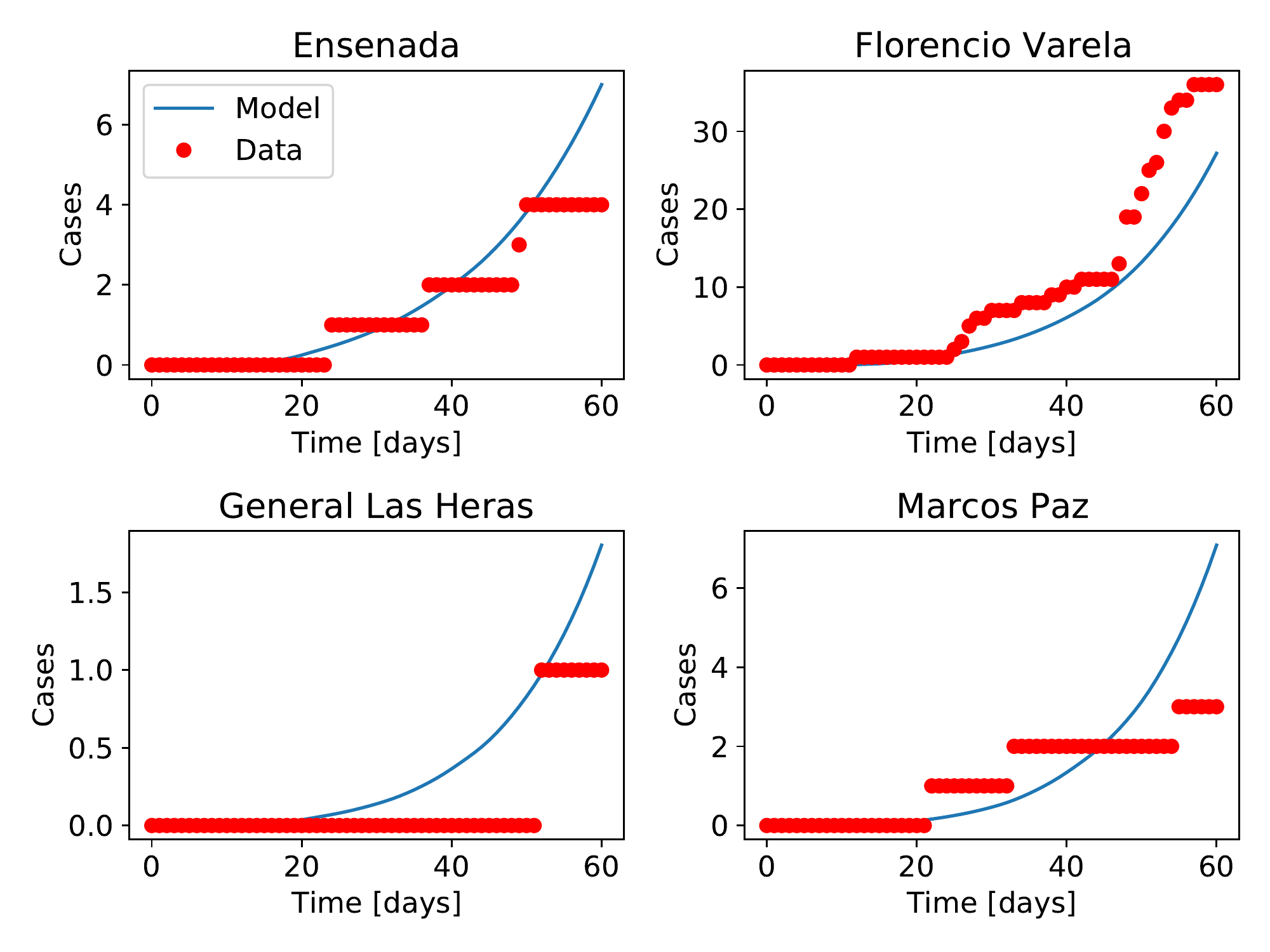}
\end{center}
\caption{Number of accumulated cases in four districts of the Buenos Aires metropolitan region that reported no infectious travelers arriving from countries with reported COVID-19 cases. Red dots indicate the official number of cases, while the blue solid lines indicate results from the deterministic coupled model.}
\label{fig:novoy}
\end{figure}

Figure \ref{fig:voyagers} shows the result of integrating the model for all the 41 districts. Results are shown for five districts in the  metropolitan area of the city of Buenos Aires, plus the city, corresponding in all cases to districts that reported cases of infected travelers coming from countries with confirmed COVID-19 infections at the early stages of the outbreak. Note that the model correctly captures the beginning of the outbreak in each district, as well as the overall evolution of the epidemic, even when districts differ in the number of cases by up to two orders of magnitude (the population does not differ that much between districts, however). Also, note that no attempt has been made to adjust parameters in the model to fit the observed data, besides using the value of $\beta_0$ obtained for the entire country.

The model also captures the time of the outbreak and the order of magnitude of the cases in districts that reported no incoming infectious travelers from countries with COVID-19. Figure \ref{fig:novoy} shows the evolution for four of such districts. Overall, this indicates that modulating transmission and coupling rates using mobility data from cell phones gives a good approximation to the dynamics of the epidemics.

\begin{figure}
\begin{center}
\includegraphics[width=.8\linewidth]{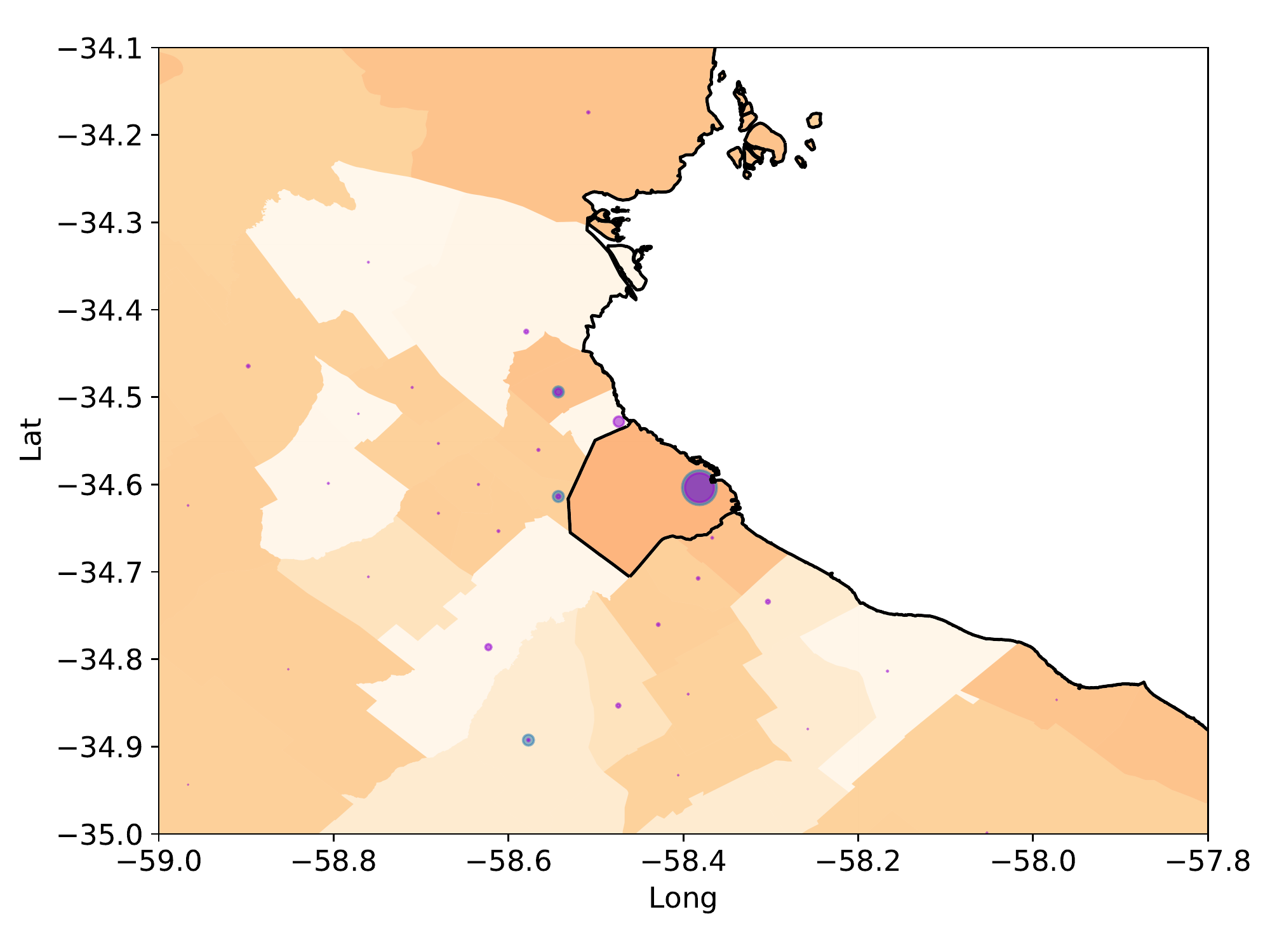}
\includegraphics[width=.8\linewidth]{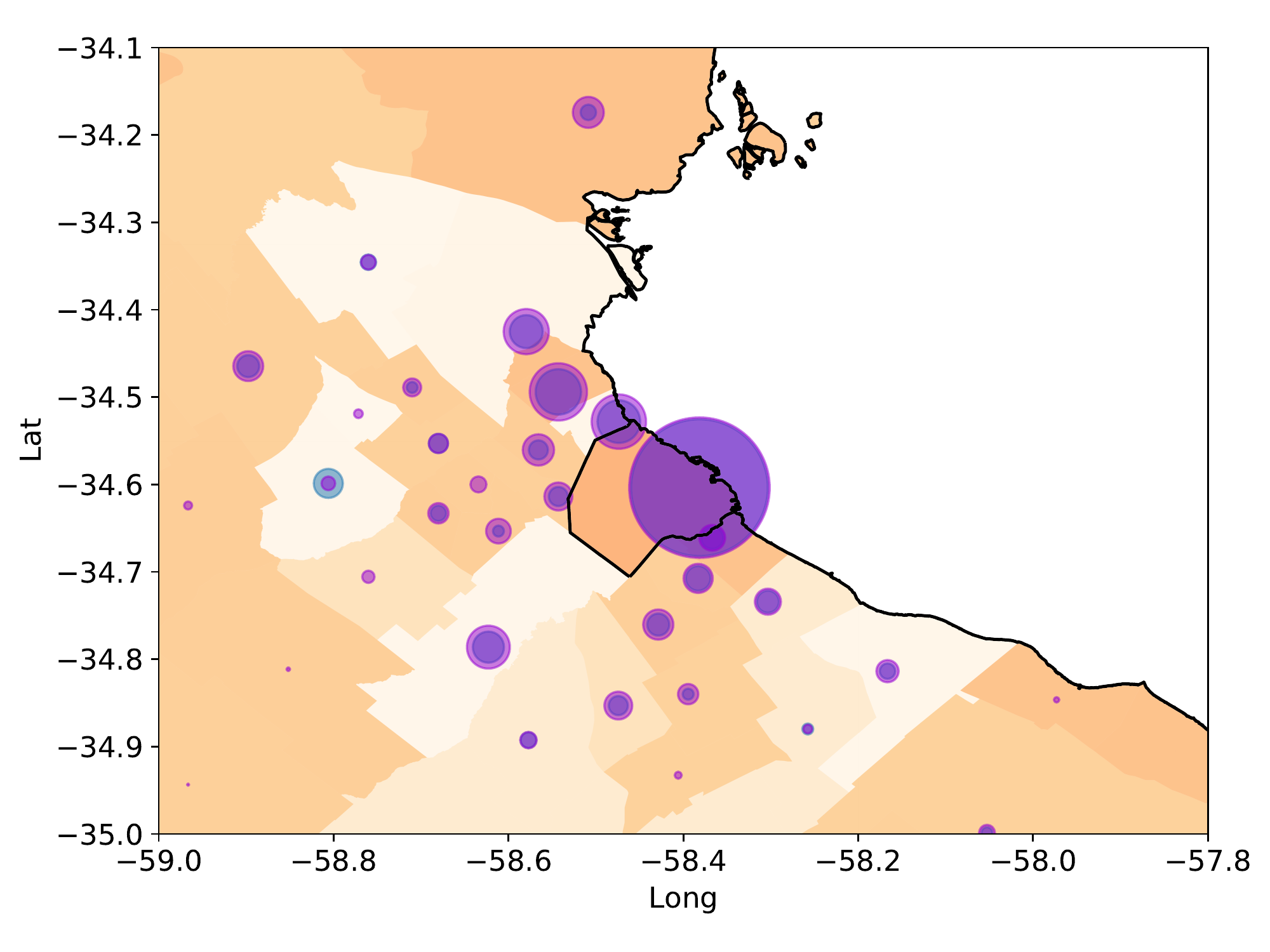}
\includegraphics[width=.8\linewidth]{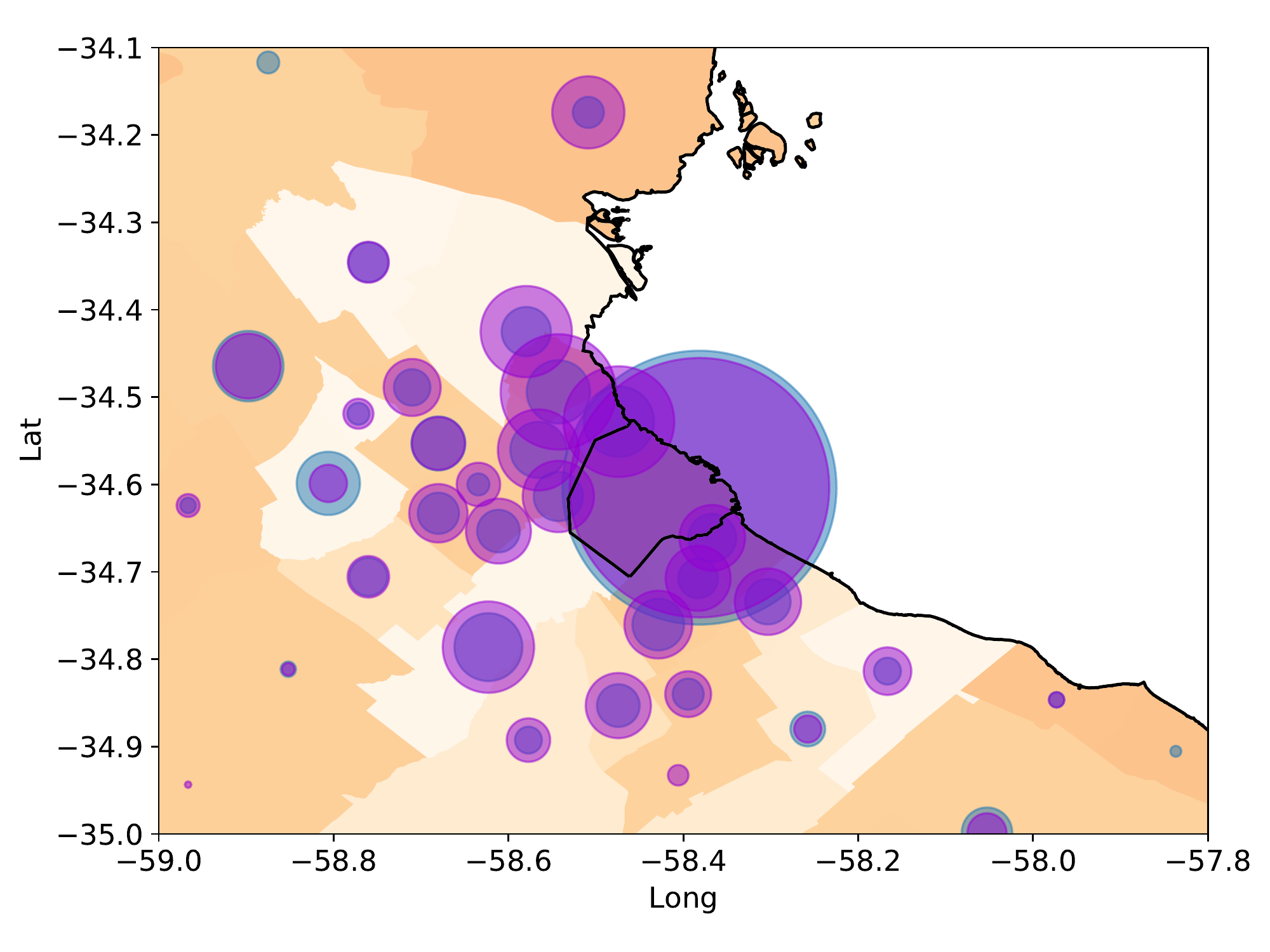}
\includegraphics[width=.8\linewidth]{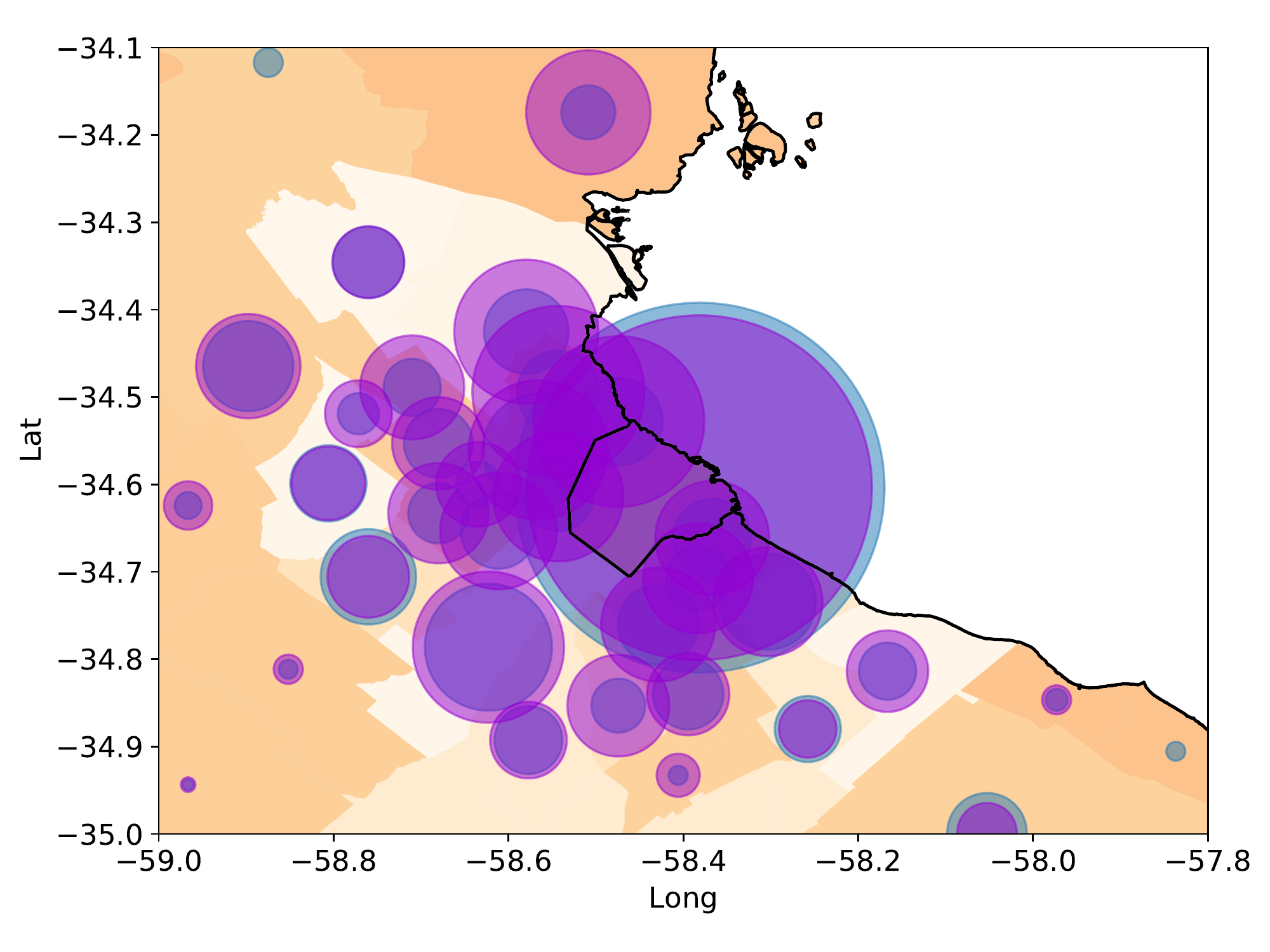}
\end{center}
\caption{Number of cases in the city of Buenos Aires (center, enclosed by black lines) and the 40 districts of the metropolitan area (in different shades). The white region corresponds to a river. From top to bottom, days 5, 20, 40, and 60 since the first reported case. The size of blue circles indicates the number of official cases, while the size of violet circles indicates the number of cases according to the deterministic coupled model.}
\label{fig:mapdet}
\end{figure}

\begin{figure}
\begin{center}
\includegraphics[width=1\linewidth]{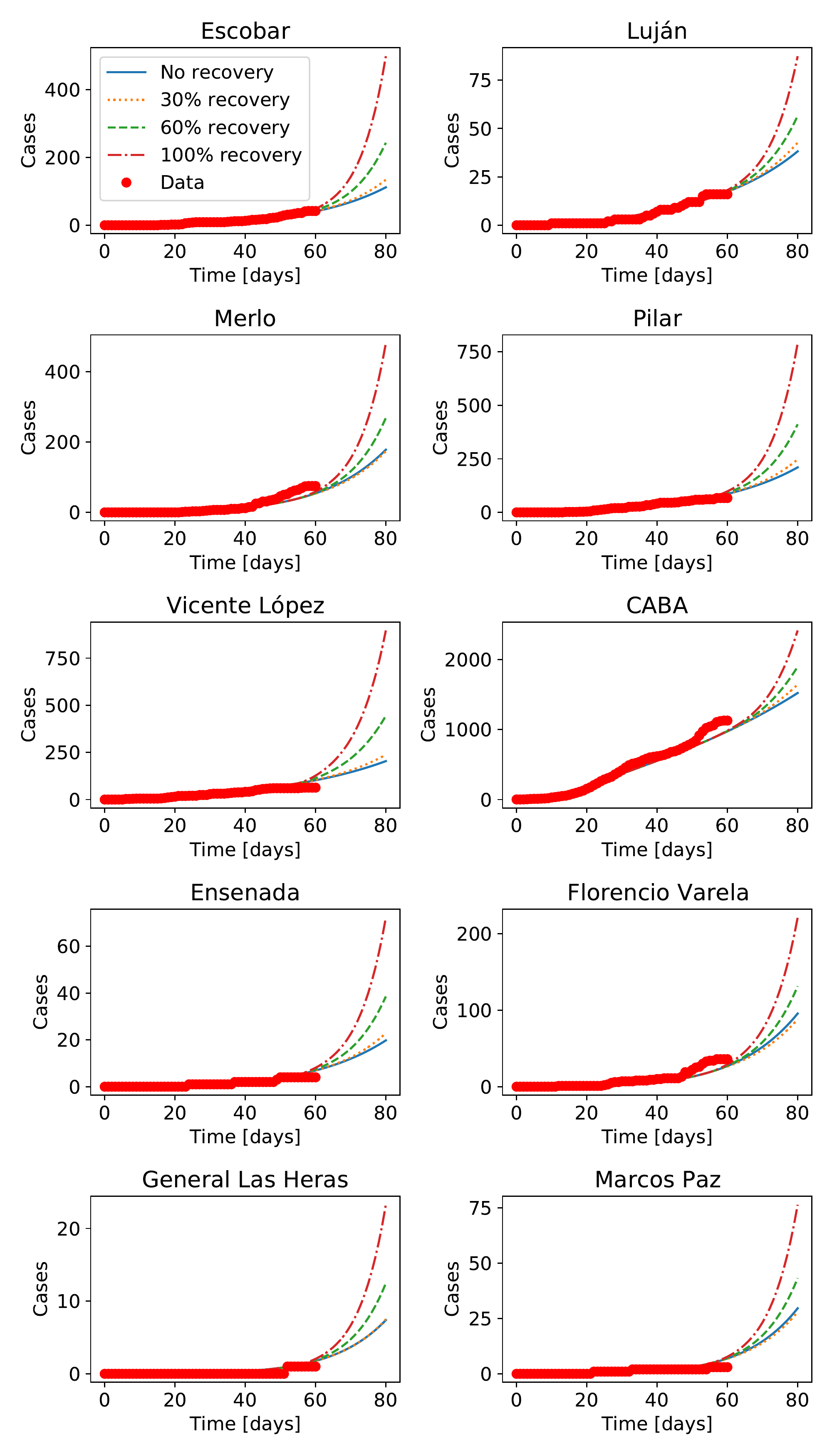}
\end{center}
\caption{Number of accumulated cases in four districts of the metropolitan region and Buenos Aires city (CABA), with forecasts for an increase in mobility recovering 30\%, 60\%, and 100\% of the values before the lockdown, starting on the last day with available data.}
\label{fig:forecast_det}
\end{figure}

Figure \ref{fig:mapdet} shows the evolution of the number of official and forecast cases for the 41 districts in a map, for days 5, 20, 40, and 60 since the first officially reported case in the country. The radius of the semi-transparent circles indicates the number of cases in each district, both for the official data as for the model output. As a result, circles with a border of a different color indicate a disparity between the model and the data: when the border is blue, the model predicts a smaller number of cases than the official reports. When the border is light violet, the model predicts an excess of cases compared with the empirical data. Note that all differences are small. Also, note that while in the first 5 days most cases were concentrated in the city of Buenos Aires, the cases spread north, west and south in time, first to closer districts in the suburbs (having more traffic and mobility to the main city), and then from these districts to others further apart. After 60 days, all districts neighbour to Buenos Aires city had a large number of cases and displayed a more or less homogeneous behavior.

Moreover, as cell phone mobility data is available daily, this type of models can be used to forecast different scenarios by increasing mobility indices to recover pre-lockdown mobility values. In this way we can analyze possible outcomes to changes in lockdown measurements. As an illustrative example, Fig.~\ref{fig:forecast_det} shows the evolution of the system in the districts in Figs.~\ref{fig:voyagers} and \ref{fig:novoy}, assuming that after the last day with available cell phone data the mobility is kept constant (i.e., the lockdown is maintained), or mobility (inside each district as well as between districts) is raised to $30\%$, $60\%$, or $100\%$ of its value before the lockdown. The more evident limitations of this model are associated with the difficulty of performing uncertainty analysis, and with the dynamics of districts with very low number of infectious individuals, where fluctuations can play a significant role. To overcome these potential limitations, we now consider the stochastic coupled model.

\subsection{Coupled stochastic models}

\begin{figure}
\begin{center}
\includegraphics[width=1\linewidth]{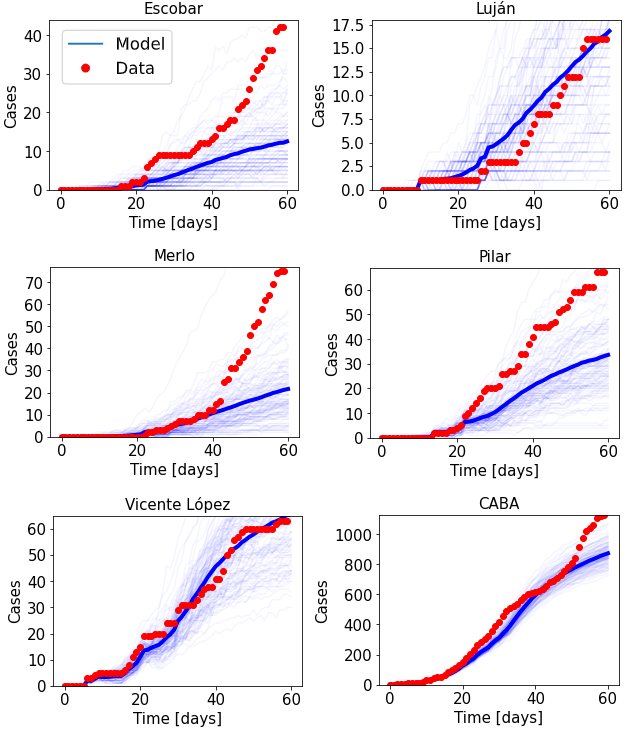}
\end{center}
\caption{Number of accumulated cases in five districts of the Buenos Aires metropolitan region, plus the city of Buenos Aires (CABA). Infectious travelers were reported in the early stages of the outbreak for all these districts. Red dots indicate the official number of cases, while blue curves indicate individual realizations of the stochastic model. The thick blue line indicates the average across all realizations.}
\label{fig:forecast_stoch_1}
\end{figure}

We now describe the results obtained from combining the stochastic model described in Eqs.~(\ref{eq:stSEIR_Si})-(\ref{eq:stSEIR_Ii}) with the mobility estimates based on the cell phone data. As in the deterministic case, we fix the initial $\beta_0$ value and introduce a smooth temporal evolution of the regional  $\beta_i$ coefficients, informed by the initial and final local mobility values. We use the empirical mixing matrix $M_{i,j}$ to determine the interaction between the exposed individuals from all pairs of regions. We report 100 simulations per region, as well as the average across all realizations of the stochastic model. As in the case of the homogeneous deterministic model described in the previous section, we note that the coupled stochastic model does not require parameter optimization and is fully determined by demographics, constants characteristic of the disease, information on infectious individuals traveling from abroad, and the mobility data.

We observe that the stochastic model can reproduce the number of cases with variable accuracy, with results that are in general compatible with the official number of cases, but occasionally with severe underestimations. Figure \ref{fig:forecast_stoch_1} shows the results for five representative districts which participated in the early stages of the outbreak. We observe that the stochastic model either provides a reasonably accurate estimate of the cumulative number of cases (e.g., for Luján, Vicente López, and CABA), or underestimates the official data. An opposite result was observed for some of the districts modeled with the coupled deterministic model (i.e., Vicente López). This difference could stem from the fact that, as opposed to the ODEs, the stochastic model can reproduce regional extinctions in the progression of the daily number of cases (unless driven by other region through the mixing matrix). It is also worth noting that, in all cases, the official data lies between different realizations of the stochastic model.

\begin{figure}
\begin{center}
\includegraphics[width=1\linewidth]{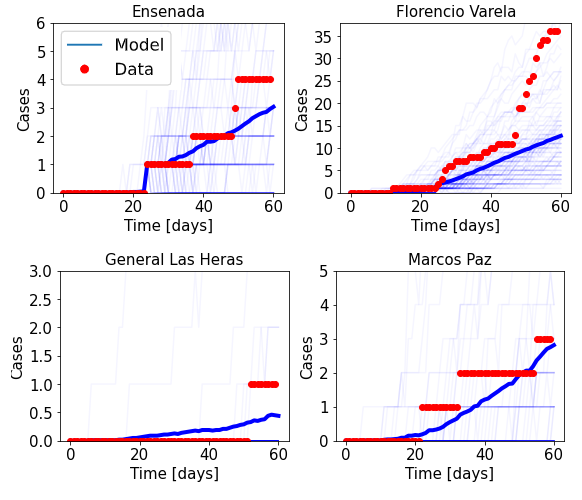}
\end{center}
\caption{Number of accumulated cases in four districts of the Buenos Aires metropolitan region, for districts that reported no infectious travelers arriving from countries with reported COVID-19 cases. Red dots indicate the official number of cases, while blue curves indicate individual realizations of the stochastic model. The thick blue line indicates the average across all realizations.}
\label{fig:forecast_stoch_2}
\end{figure}

Again, note that the stochastic model captured the time of the outbreak, and the order  of  magnitude  of  the  cases  in  districts  that  reported  no incoming infectious travelers from countries with COVID-19 cases (Figure \ref{fig:forecast_stoch_2}). Note that the stochastic model applied to this situation presents the strength of factoring in the non-negligible possibility of focus extinction, and thus can result in forecasts following more closely the empirical number of cases compared to the homogeneous model. 

Both the stochastic and homogeneous models are driven by empirical mobility data; however, the compulsory lockdown (starting on March 19th) resulted in a drastic drop of both local and inter-regional mobility. We asked whether such overall change in mobility sufficed to explain the goodness of fit of our models, or whether the finer structure of the mixing matrix and the local mobility estimates also played a significant role. For this purpose, we compared the results against a null model obtained from repeating the above presented simulations across 1000 independent realizations, randomly scrambling the ordering of the regions in the mixing matrix $M_{i,j}$ and in the infection rate vector $\beta_i$. Note that both variables were scrambled only once per simulation, i.e., not once per time step. Finally, we derived a p-value by counting the number of times the goodness of fit (computed as the inverse of the mean squared error [MSE]) of the model with shuffled $M_{i,j}$ and $\beta_i$ was better than the one obtained from the unshuffled data. We conclude that the structure of the empirical mixing matrix plays a significant role in the goodness of fit of our predictions, since the average MSE is higher for the shuffled than for the unshuffled mixing matrix ($p<0.001$, and $p<0.05$ for 17 of the regions). Conversely, we did not find a significant effect of the local mobility. This is consistent with cursory observations that suggest higher heterogeneity in the inter-regional mobility compared to the local circulation.

\section{Conclusions \label{sec:conclusions}}

In this work we presented results of different approaches to model the evolution of the COVID-19 epidemic in Argentina, with a special focus on the megacity comprised by the city of Buenos Aires and its large metropolitan area. We progressively moved from models that assume homogeneous mixing of the population, towards more complex models that allow for inhomogeneities between individual districts. To inform these models, we incorporated the number of infectious individuals arriving from other countries with confirmed COVID-19 cases to our models, as well as cell phone data to estimate mobility within and between districts.

First, the effects of some proposed NPI policies and the consequences of rapid homogeneization of the epidemics were considered. We conclude that it is unrealistic to implement certain policies, e.g., alternating between lockdowns and periods of mobility. We also note that without empirical mobility estimates, performing ensemble averages varying unknown parameters over multiple realizations of coupled models results in an evolution compatible with that of homogeneous deterministic models, thus providing very little extra information.

It is thus apparent that in highly populated regions, the evolution  of the  epidemics  can  be  modelled  to  a  certain  degree  by  locally homogeneous models, and by the coupling of these models between regions that display  inhomogeneous mixing between themselves. We showed that mixing and contact matrices estimated from cell phone data could be used to obtain reasonable estimations of the infection rates, both for stochastic and for deterministic models. The predictions of both types of models compared favourably to the official data during the first 60 days of the outbreak. Crucially, this performance depended on the structure of the mixing matrix and not only on its post-lockdown modulation, as shown by the application of permutation tests.  

Direct data analysis, as well as comparisons with the results from the models, indicate that while in the first 5 days most cases were concentrated in the city of Buenos Aires, the disease spread north, west and south as a function of time. The spread occurred first to closer districts in the suburbs, and from  these  districts to other  further apart in the region. The models captured this spatio-temporal evolution, including the time of the first cases in each district, and the order of magnitude of the number of cases in a situation where the number of infectious individuals varied significantly from district to district.

As in the introduction section, we conclude this work with a word of caution. We remind the reader again that simple epidemiological models, even when fed with data stemming from population census and detailed cell phone mobility estimates, are limited in what they can tell us. These limitations arise both from inefficiencies at the core of these models, as well as from shortcomings of the available data. Making these models more complex does not necessarily improve their forecasting capabilities, specially in a situation in which solutions (and small differences) grow, in some cases, exponentially in time. As a result, these models should only be considered as ways of exploring possible solutions of the system, and extreme caution should be exercised whenever these models are used to generate advice in the process of public policy decision making.

\section*{Acknowledgements}
E.T, P.B, G.B.M., and P.D.M. wish to acknowledge enlightening discussions with colleagues with whom they had the pleasure to share a month of work in a first response modeling committee, while learning these lessons from being challenged by COVID-19: Hern\'an Solari, Marcelo Kuperman, Gustavo Sibona, Zulma Ortiz, Gabriel Fabricius, Juan Aparicio, Veronica Simoy, and Ignacio Simoy (who also provided invaluable help in the processing of the data). The authors also acknowledge Sebasti\'an Pinto, Facundo Fainstein, Sebastian Geli, Martín Coll, Federico Albanese, Franco Castellacci, and Sof\'{\i}a del Pozo for providing support in data analysis and visualization. E.T. acknowledges the support of the Toyoko 2020 Cloud Credits for Science program, and thanks Sebastián Bassi and Virginia González for their continuous support. 
\bibliography{ms}

\begin{thebibliography}{10}
\expandafter\ifx\csname url\endcsname\relax
  \def\url#1{\texttt{#1}}\fi
\expandafter\ifx\csname urlprefix\endcsname\relax\def\urlprefix{URL }\fi
\expandafter\ifx\csname href\endcsname\relax
  \def\href#1#2{#2} \def\path#1{#1}\fi

\bibitem{EarlyLombardia}
D.~Cereda, M.~Tirani, F.~Rovida, V.~Demicheli, M.~Ajelli, P.~Poletti,
  F.~Trentini, G.~Guzzetta, V.~Marziano, A.~Barone, M.~Magoni, S.~Deandrea,
  G.~Diurno, M.~Lombardo, M.~Faccini, A.~Pan, R.~Bruno, E.~Pariani,
  G.~Grasselli, A.~Piatti, M.~Gramegna, F.~Baldanti, A.~Melegaro, S.~Merler,
  \href{https://arxiv.org/abs/2003.09320}{The early phase of the {COVID}-19
  outbreak in {L}ombardy, {I}taly} (2020).
\newblock \href {http://arxiv.org/abs/2003.09320} {\path{arXiv:2003.09320}}.
\newline\urlprefix\url{https://arxiv.org/abs/2003.09320}

\bibitem{DiamondPrincess}
T.~W. Russell, J.~Hellewell, C.~I. Jarvis, K.~van Zandvoort, S.~Abbott,
  R.~Ratnayake, , S.~Flasche, R.~M. Eggo, A.~J. Kucharski,
  \href{https://doi.org/10.2807/1560-7917.ES.2020.25.12.2000256}{Estimating the
  infection and case fatality ratio for coronavirus disease ({COVID}-19) using
  age-adjusted data from the outbreak on the {D}iamond {P}rincess cruise ship,
  {F}ebruary 2020}, Euro Surveillance 25~(12) (2020) 2000256.
\newline\urlprefix\url{https://doi.org/10.2807/1560-7917.ES.2020.25.12.2000256}

\bibitem{Ferguson2006}
N.~M. Ferguson, D.~A.~T. Cummings, C.~Fraser, J.~C. Cajka, P.~C. Cooley, D.~S.
  Burke, \href{https://doi.org/10.1038/nature04795}{Strategies for mitigating
  an influenza pandemic}, Nature 442 (2006) 448--452.
\newblock \href {https://doi.org/10.1038/nature04795}
  {\path{doi:10.1038/nature04795}}.
\newline\urlprefix\url{https://doi.org/10.1038/nature04795}

\bibitem{Halloran2008}
M.~E. Halloran, N.~M. Ferguson, S.~Eubank, I.~M. Longini, D.~A.~T. Cummings,
  B.~Lewis, S.~Xu, C.~Fraser, A.~Vullikanti, T.~C. Germann, D.~Wagener,
  R.~Beckman, K.~Kadau, C.~Barrett, C.~A. Macken, D.~S. Burke, P.~Cooley,
  \href{https://www.pnas.org/content/105/12/4639}{Modeling targeted layered
  containment of an influenza pandemic in the {U}nited {S}tates}, Proceedings
  of the National Academy of Sciences 105~(12) (2008) 4639--4644.
\newblock \href
  {http://arxiv.org/abs/https://www.pnas.org/content/105/12/4639.full.pdf}
  {\path{arXiv:https://www.pnas.org/content/105/12/4639.full.pdf}}, \href
  {https://doi.org/10.1073/pnas.0706849105}
  {\path{doi:10.1073/pnas.0706849105}}.
\newline\urlprefix\url{https://www.pnas.org/content/105/12/4639}

\bibitem{VespignaniGleam2010}
D.~Balcan, B.~Gonçalves, H.~Hu, J.~J. Ramasco, V.~Colizza, A.~Vespignani,
  \href{http://www.sciencedirect.com/science/article/pii/S1877750310000438}{Modeling
  the spatial spread of infectious diseases: {T}he {GL}obal {E}pidemic and
  {M}obility computational model}, Journal of Computational Science 1~(3)
  (2010) 132 -- 145.
\newblock \href {https://doi.org/https://doi.org/10.1016/j.jocs.2010.07.002}
  {\path{doi:https://doi.org/10.1016/j.jocs.2010.07.002}}.
\newline\urlprefix\url{http://www.sciencedirect.com/science/article/pii/S1877750310000438}

\bibitem{Arenas2018}
D.~Soriano-Pa\~nos, L.~Lotero, A.~Arenas, J.~G\'omez-Garde\~nes,
  \href{https://link.aps.org/doi/10.1103/PhysRevX.8.031039}{Spreading processes
  in multiplex metapopulations containing different mobility networks},
  Physical Review X 8 (2018) 031039.
\newblock \href {https://doi.org/10.1103/PhysRevX.8.031039}
  {\path{doi:10.1103/PhysRevX.8.031039}}.
\newline\urlprefix\url{https://link.aps.org/doi/10.1103/PhysRevX.8.031039}

\bibitem{VespingnaniFebrero2020}
D.~Mistry, M.~Litvinova, A.~{Pastore y Piontti}, M.~Chinazzi, L.~Fumanelli,
  M.~F.~C. Gomes, S.~A. Haque, Q.-H. Liu, K.~Mu, X.~Xiong, M.~E. Halloran,
  I.~M. {Longini Jr.}, S.~Merler, M.~Ajelli, A.~Vespignani, Inferring
  high-resolution human mixing patterns for disease modeling (2020).
\newblock \href {http://arxiv.org/abs/2003.01214} {\path{arXiv:2003.01214}}.

\bibitem{YMoreno2018}
S.~Arregui, A.~Aleta, J.~Sanz, Y.~Moreno,
  \href{https://doi.org/10.1371/journal.pcbi.1006638}{Projecting social contact
  matrices to different demographic structures}, PLOS Computational Biology
  14~(12) (2018) 1--18.
\newblock \href {https://doi.org/10.1371/journal.pcbi.1006638}
  {\path{doi:10.1371/journal.pcbi.1006638}}.
\newline\urlprefix\url{https://doi.org/10.1371/journal.pcbi.1006638}

\bibitem{Kuniya2020}
T.~Kuniya, \href{http://dx.doi.org/10.3390/jcm9030789}{Prediction of the
  epidemic peak of coronavirus disease in {J}apan, 2020}, Journal of Clinical
  Medicine 9~(3) (2020) 789.
\newblock \href {https://doi.org/10.3390/jcm9030789}
  {\path{doi:10.3390/jcm9030789}}.
\newline\urlprefix\url{http://dx.doi.org/10.3390/jcm9030789}

\bibitem{Brockmann2020}
B.~F. Maier, D.~Brockmann,
  \href{https://science.sciencemag.org/content/early/2020/04/07/science.abb4557}{Effective
  containment explains subexponential growth in recent confirmed {COVID}-19
  cases in {C}hina}, Science (2020).
\newblock \href
  {http://arxiv.org/abs/https://science.sciencemag.org/content/early/2020/04/07/science.abb4557.full.pdf}
  {\path{arXiv:https://science.sciencemag.org/content/early/2020/04/07/science.abb4557.full.pdf}},
  \href {https://doi.org/10.1126/science.abb4557}
  {\path{doi:10.1126/science.abb4557}}.
\newline\urlprefix\url{https://science.sciencemag.org/content/early/2020/04/07/science.abb4557}

\bibitem{Chile2020}
A.~Cancin, C.~Castillo, P.~Gajardo, R.~Lecaros, C.~Muñoz, C.~Naranjo,
  J.~Ortega, H.~Ramirez, Report 2: {E}stimation of maximal {ICU} beds demand
  for {C}ovid-19 outbreak in {S}antiago, {C}hile, Tech. rep. (2020).
\newblock \href
  {http://arxiv.org/abs/http://www.cmm.uchile.cl/wp-content/uploads/2020/03/Reporte2_CMM_AM2V_CEPS.pdf}
  {\path{arXiv:http://www.cmm.uchile.cl/wp-content/uploads/2020/03/Reporte2_CMM_AM2V_CEPS.pdf}}.

\bibitem{Arenas2020}
A.~Arenas, W.~Cota, J.~Gomez-Gardenes, S.~G{\'o}mez, C.~Granell, J.~T.
  Matamalas, D.~Soriano-Panos, B.~Steinegger,
  \href{https://www.medrxiv.org/content/early/2020/03/23/2020.03.21.20040022}{A
  mathematical model for the spatiotemporal epidemic spreading of {COVID}19},
  medRxiv (2020).
\newblock \href
  {http://arxiv.org/abs/https://www.medrxiv.org/content/early/2020/03/23/2020.03.21.20040022.full.pdf}
  {\path{arXiv:https://www.medrxiv.org/content/early/2020/03/23/2020.03.21.20040022.full.pdf}},
  \href {https://doi.org/10.1101/2020.03.21.20040022}
  {\path{doi:10.1101/2020.03.21.20040022}}.
\newline\urlprefix\url{https://www.medrxiv.org/content/early/2020/03/23/2020.03.21.20040022}

\bibitem{Ferguson1_2020}
N.~M. Ferguson, D.~Laydon, G.~Nedjati-Gilani, et~al., Report 9: Impact of
  non-pharmaceutical interventions ({NPIs}) to reduce {COVID}-19 mortality and
  healthcare demand, Tech. rep. (2020).
\newblock \href {http://arxiv.org/abs/https://doi.org/10.25561/77482.}
  {\path{arXiv:https://doi.org/10.25561/77482.}}, \href
  {https://doi.org/10.25561/77482} {\path{doi:10.25561/77482}}.

\bibitem{UriAlon2020}
O.~Karin, Y.~M. Bar-On, T.~Milo, I.~Katzir, A.~Mayo, Y.~Korem, B.~Dudovich,
  E.~Yashiv, A.~J. Zehavi, N.~Davidovich, R.~Milo, U.~Alon,
  \href{https://www.medrxiv.org/content/early/2020/04/28/2020.04.04.20053579}{Adaptive
  cyclic exit strategies from lockdown to suppress {COVID}-19 and allow
  economic activity}, medRxiv (2020).
\newblock \href
  {http://arxiv.org/abs/https://www.medrxiv.org/content/early/2020/04/28/2020.04.04.20053579.full.pdf}
  {\path{arXiv:https://www.medrxiv.org/content/early/2020/04/28/2020.04.04.20053579.full.pdf}},
  \href {https://doi.org/10.1101/2020.04.04.20053579}
  {\path{doi:10.1101/2020.04.04.20053579}}.
\newline\urlprefix\url{https://www.medrxiv.org/content/early/2020/04/28/2020.04.04.20053579}

\bibitem{YMorenoReport2020}
D.~Martin-Calvo, A.~Aleta, A.~Pentland, Y.~Moreno, E.~Moro, Effectiveness of
  social distancing strategies for protecting a community from a pandemic with
  a data-driven contact network based on census and real-world mobility data,
  Tech. rep. (2020).
\newblock \href
  {http://arxiv.org/abs/https://covid-19-sds.github.io/assets/pdfs/Preliminary_Report_Effectiveness_of_social_distance_strategies_COVID-19.pdf}
  {\path{arXiv:https://covid-19-sds.github.io/assets/pdfs/Preliminary_Report_Effectiveness_of_social_distance_strategies_COVID-19.pdf}}.

\bibitem{YMorenoVespignaniReport2020}
A.~Aleta, D.~Martin-Corral, A.~Pastore~y Piontti, M.~Ajelli, M.~Litvinova,
  N.~E. Dean, M.~E. Halloran, I.~M. Longini, S.~Merler, A.~Pentland,
  A.~Vespignani, E.~Moro, Y.~Moreno, Modeling the impact of social distancing,
  testing, contact tracing and household quarantine on second-wave scenarios of
  the {COVID}-19 epidemic, Tech. rep. (2020).
\newblock \href
  {http://arxiv.org/abs/https://www.mobs-lab.org/uploads/6/7/8/7/6787877/tracing_main_may4.pdf}
  {\path{arXiv:https://www.mobs-lab.org/uploads/6/7/8/7/6787877/tracing_main_may4.pdf}}.

\bibitem{Prem2020}
K.~Prem, Y.~Liu, T.~W. Russell, A.~J. Kucharski, R.~M. Eggo, N.~Davies,
  S.~Flasche, S.~Clifford, C.~A.~B. Pearson, J.~D. Munday, S.~Abbott, H.~Gibbs,
  A.~Rosello, B.~J. Quilty, T.~Jombart, F.~Sun, C.~Diamond, A.~Gimma, K.~[van
  Zandvoort], S.~Funk, C.~I. Jarvis, W.~J. Edmunds, N.~I. Bosse, J.~Hellewell,
  M.~Jit, P.~Klepac,
  \href{http://www.sciencedirect.com/science/article/pii/S2468266720300736}{The
  effect of control strategies to reduce social mixing on outcomes of the
  {COVID}-19 epidemic in {W}uhan, {C}hina: a modelling study}, The Lancet
  Public Health 5~(5) (2020) e261 -- e270.
\newblock \href {https://doi.org/https://doi.org/10.1016/S2468-2667(20)30073-6}
  {\path{doi:https://doi.org/10.1016/S2468-2667(20)30073-6}}.
\newline\urlprefix\url{http://www.sciencedirect.com/science/article/pii/S2468266720300736}

\bibitem{Giordano2020}
G.~Giordano, F.~Blanchini, R.~Bruno, P.~Colaneri, A.~Di~Filippo, A.~Di~Matteo,
  M.~Colaneri, Modelling the {COVID}-19 epidemic and implementation of
  population-wide interventions in {I}taly, Nature Medicine (2020).
\newblock \href {https://doi.org/https://doi.org/10.1038/s41591-020-0883-7}
  {\path{doi:https://doi.org/10.1038/s41591-020-0883-7}}.

\bibitem{VespignaniRt2018}
Q.-H. Liu, M.~Ajelli, A.~Aleta, S.~Merler, Y.~Moreno, A.~Vespignani,
  \href{https://www.pnas.org/content/115/50/12680}{Measurability of the
  epidemic reproduction number in data-driven contact networks}, Proceedings of
  the National Academy of Sciences 115~(50) (2018) 12680--12685.
\newblock \href
  {http://arxiv.org/abs/https://www.pnas.org/content/115/50/12680.full.pdf}
  {\path{arXiv:https://www.pnas.org/content/115/50/12680.full.pdf}}, \href
  {https://doi.org/10.1073/pnas.1811115115}
  {\path{doi:10.1073/pnas.1811115115}}.
\newline\urlprefix\url{https://www.pnas.org/content/115/50/12680}

\bibitem{Li2020}
R.~Li, S.~Pei, B.~Chen, Y.~Song, T.~Zhang, W.~Yang, J.~Shaman,
  \href{https://science.sciencemag.org/content/368/6490/489}{Substantial
  undocumented infection facilitates the rapid dissemination of novel
  coronavirus ({SARS-CoV-2})}, Science 368~(6490) (2020) 489--493.
\newblock \href
  {http://arxiv.org/abs/https://science.sciencemag.org/content/368/6490/489.full.pdf}
  {\path{arXiv:https://science.sciencemag.org/content/368/6490/489.full.pdf}},
  \href {https://doi.org/10.1126/science.abb3221}
  {\path{doi:10.1126/science.abb3221}}.
\newline\urlprefix\url{https://science.sciencemag.org/content/368/6490/489}

\bibitem{Zhang2020}
J.~Zhang, M.~Litvinova, W.~Wang, Y.~Wang, X.~Deng, X.~Chen, M.~Li, W.~Zheng,
  L.~Yi, X.~Chen, Q.~Wu, Y.~Liang, X.~Wang, J.~Yang, K.~Sun, I.~M. Longini,
  M.~E. Halloran, P.~Wu, B.~J. Cowling, S.~Merler, C.~Viboud, A.~Vespignani,
  M.~Ajelli, H.~Yu,
  \href{http://www.sciencedirect.com/science/article/pii/S1473309920302309}{Evolving
  epidemiology and transmission dynamics of coronavirus disease 2019 outside
  {H}ubei province, {C}hina: a descriptive and modelling study}, The Lancet
  Infectious Diseases (2020).
\newblock \href {https://doi.org/https://doi.org/10.1016/S1473-3099(20)30230-9}
  {\path{doi:https://doi.org/10.1016/S1473-3099(20)30230-9}}.
\newline\urlprefix\url{http://www.sciencedirect.com/science/article/pii/S1473309920302309}

\bibitem{Christakis2020}
J.~S. Jia, X.~Lu, Y.~Yuan, G.~Xu, J.~Jia, N.~A. Christakis,
  \href{https://doi.org/10.1038/s41586-020-2284-y}{Population flow drives
  spatio-temporal distribution of {COVID}-19 in {C}hina}, Nature (2020).
\newblock \href
  {http://arxiv.org/abs/https://doi.org/10.1038/s41586-020-2284-y}
  {\path{arXiv:https://doi.org/10.1038/s41586-020-2284-y}}, \href
  {https://doi.org/10.1038/s41586-020-2284-y}
  {\path{doi:10.1038/s41586-020-2284-y}}.
\newline\urlprefix\url{https://doi.org/10.1038/s41586-020-2284-y}

\bibitem{Lekone}
P.~E. Lekone, B.~F. F., Statistical inference in a stochastic epidemic {SEIR}
  model with control intervention: {E}bola as a case study, Biometrics 62
  (2006) 1170–1177.

\bibitem{Feng}
Z.~Feng, Final and peak epidemic sizes for {SEIR} models with quarantine and
  isolation, Mathematical Biosciences and Engineering 4 (2007) 675.

\bibitem{GHolmes}
J.~Guckenheimer, P.~Holmes, Nonlinear Oscillations, Dynamical Systems, and
  Bifurcations of Vector Fields, Springer-Verlag, 1983.

\bibitem{Ferguson2_2020}
S.~Flaxman, S.~Mishra, A.~Gandy, et~al, Estimating the number of infections and
  the impact of non-pharmaceutical interventions on {COVID}-19 in 11 {E}uropean
  countries, Tech. rep. (2020).
\newblock \href {http://arxiv.org/abs/https://doi.org/10.25561/77731}
  {\path{arXiv:https://doi.org/10.25561/77731}}, \href
  {https://doi.org/10.25561/77731} {\path{doi:10.25561/77731}}.

\bibitem{Anapolsky2014}
C.~Sarraute, N.~Ponieman, C.~Lang, S.~Anapolsky,
  \href{https://netmob.org/www15/assets/img/netmob15_book_of_abstracts_posters.pdf}{The
  city pulse of buenos aires}, in: NetMob 2015 (Fourth Conference on the
  Scientific Analysis of Mobile Phone Datasets), MIT Media Lab, Cambridge, USA,
  8-10 April 2015, 2015, pp. 54--56.
\newline\urlprefix\url{https://netmob.org/www15/assets/img/netmob15_book_of_abstracts_posters.pdf}

\bibitem{MinSal}
{Ministerio de Salud de la Naci\'on de la Rep\'ublica Argentina}, Official data
  for {COVID}-19 cases in {A}rgentina,
  \url{https://www.argentina.gob.ar/salud/coronavirus-COVID-19/sala-situacion}
  (May 2020).

\bibitem{Lloyd}
A.~L. Lloyd, R.~M. May, Spatial heterogeneity in epidemic models, J. Theor.
  Biol. 179 (1996) 1–11.

\bibitem{blondel2008fast}
V.~D. Blondel, J.-L. Guillaume, R.~Lambiotte, E.~Lefebvre, Fast unfolding of
  communities in large networks, Journal of statistical mechanics: theory and
  experiment 2008~(10) (2008) P10008.

\bibitem{hu2005efficient}
Y.~Hu, Efficient, high-quality force-directed graph drawing, Mathematica
  Journal 10~(1) (2005) 37--71.

\end{thebibliography}

\end{document}